\documentclass[sigconf, screen, pageno, nonacm]{acmart}
\usepackage[]{hyperref}
\usepackage{subcaption}
\usepackage{multirow}

\usepackage{amssymb}
\usepackage{tikz}
\usepackage{amsmath}
\usepackage{xspace}
\usepackage{natbib}
\usepackage{enumitem}
\usepackage[ruled,vlined, linesnumbered, noend]{algorithm2e}
\usepackage{algorithmic}
\usepackage{pifont}
\usepackage{xfrac}
\usepackage{balance}
\usepackage{amsthm}
\usepackage{caption}
\usepackage{graphicx}
\usepackage{booktabs}

\begin{document}
\sloppy{
\emergencystretch 3em
}

\newcommand{\iid}{\texttt{iid}\xspace}
\newcommand{\noniid}{non-\texttt{iid}\xspace}
\newcommand{\systemName}{\texttt{PowerTrip}\xspace}
\newcommand{\cselect}{\texttt{CSelect}\xspace}
\newcommand{\cprov}{\texttt{CProv}\xspace}
\newcommand{\cscale}{\texttt{CScale}\xspace}
\newcommand{\emissionunit}{\ensuremath{g\cdot CO_{2}eq}\xspace}
\newcommand{\ciunit}{\ensuremath{g\cdot CO_{2}eq/kWh}\xspace}
\newcommand{\cciunit}{\ensuremath{g\cdot CO_{2}eq/cycle}\xspace}
\newcommand{\efunit}{\ensuremath{cycle/kWh}\xspace}
\newcommand{\xmark}{\ding{55}}%
\newcommand{\tickmark}{\ding{51}}%

\date{}

\title[PowerTrip]{PowerTrip: Exploiting Federated Heterogeneous Datacenter Power for Distributed ML Training
}

\author{Talha Mehboob}
\affiliation{
  \institution{University of Massachusetts Amherst}
    \city{}
  \country{}
}
\email{tmehboob@umass.edu}

\author{Luanzheng Guo}
\affiliation{
  \institution{Pacific Northwest National Lab}
    \city{}
  \country{}
}
\email{lenny.guo@pnnl.gov}

\author{Nathan Tallent}
\affiliation{
  \institution{Pacific Northwest National Lab}
    \city{}
  \country{}
}
\email{nathan.tallent@pnnl.gov}

\author{Michael Zink}
\affiliation{
  \institution{University of Massachusetts Amherst}
    \city{}
  \country{}
}
\email{mzink@cas.umass.edu}

\author{David Irwin}
\affiliation{
  \institution{University of Massachusetts Amherst}
    \city{}
  \country{}
}
\email{irwin@ecs.umass.edu}

\begin{abstract}
The exponential growth of large-scale AI models has led to computational and power demands that can exceed the capacity of a single data center.  This is due to the limited power supplied by regional grids that leads to limited regional computational power. Consequently, distributing training workloads across geographically distributed sites has become essential. However, this approach introduces a significant challenge in the form of communication overhead, creating a fundamental trade-off between the performance gains from accessing greater aggregate power and the performance losses from increased network latency. Although prior work has focused on reducing communication volume or using heuristics for distribution, these methods assume constant homogeneous power supplies and ignore the challenge of heterogeneous power availability between sites.

To address the challenge of training large models in power-constrained, geo-distributed environments, we introduce \systemName, a system that dynamically selects a subset of sites during runtime to optimize the power-communication trade-off. Specifically, \systemName selects sites based on a power-to-cost heuristic, prioritizing those with high power availability and low network latency. \systemName employs a dynamic greedy approach and uses the marginal gain in training efficiency, i.e., accuracy improvement per unit of time, to optimize for the number of sites where the performance penalty from network overhead negates the benefit of adding more computational power. Our evaluation, which uses real-world Google power traces to model realistic power capacity constraints, demonstrates that \systemName can reduce time-to-accuracy by up to 50\% compared to existing baseline policies. 
\end{abstract}

\maketitle 

\section{Introduction}
\label{sec:intro}

\begin{figure}[t]
    \centering
    \includegraphics[width=\linewidth]{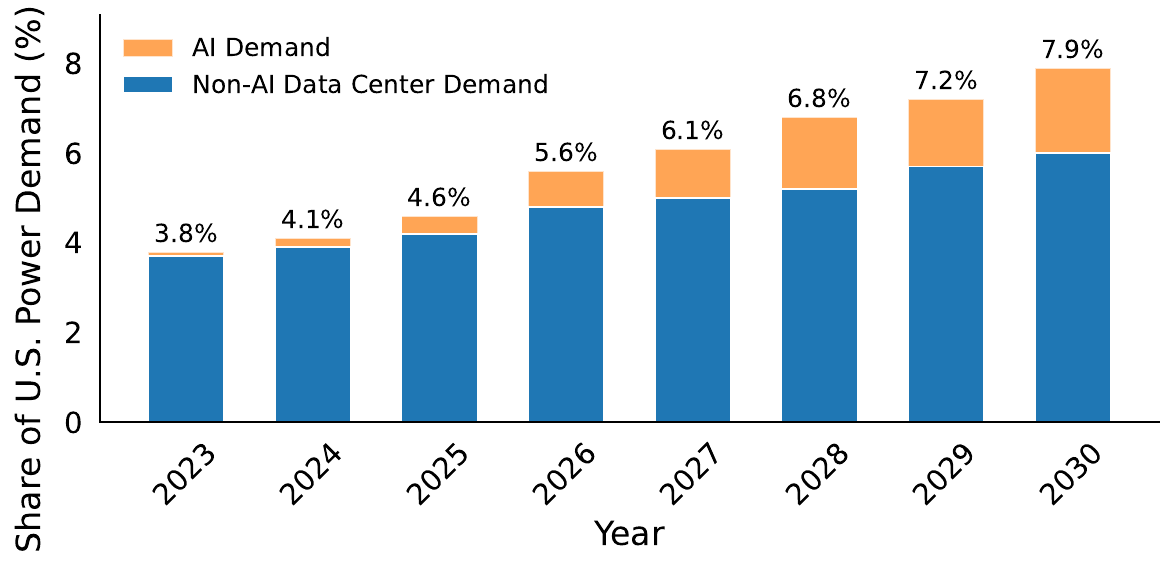}
    \vspace{-0.6cm}
    \caption{\textbf{\emph{Projected growth of data center power demand~\cite{aidemandgrowth}.}}}
    \label{fig:AI_demand_growth}
    \vspace{-0.7cm}
\end{figure}

The rapid expansion of generative AI models, such as GPT~\cite{achiam2023gpt}, OPT~\cite{zhang2022opt}, and LLaMA~\cite{touvron2023llama}, has dramatically increased the computational and energy demands of AI training workloads~\cite{kaplan2020scaling, jiang2024megascale}. By 2030, data centers are projected to consume approximately 8\% of total U.S. electricity. AI is the primary driver of this growth, with its share of data center power demand expected to increase from just 2.7\% in 2023 to over 30\% by year 2030~\cite{aidemandgrowth} (see \autoref{fig:AI_demand_growth}). Some analyses suggest this figure could be as high as 50-70\% by then~\cite{power2024expanding, lee2025breaking}.
This exponential increase in demand will soon reach a fundamental physical constraint: the limited capacity of regional power grids. While AI workloads are scaling rapidly, the underlying grid infrastructure remains stagnant, resulting in a situation where the power demand from a single data center can exceed the ability of the local grid to reliably supply power to it~\cite{avelar2023ai, norris2025rethinking}. In addition to limited power availability, data centers may also have limited compute resources due to GPU shortages within a single region~\cite{yang2023skypilot}.
The limited availability of both power and compute resources can significantly degrade the performance of ML training in the data centers.

Power and computational resource constraints make the distribution of training workloads across multiple geographically distributed sites a necessity. Prior work has explored distributing ML training for large models across multiple geographical locations to leverage additional computational resources and improve training efficiency~\cite{tang2024fusionllm, ye2024openfedllm, dong2025beyond}. Decentralized~\cite{strati2024ml, shi2024greenllm, tang2024fusionllm} and federated~\cite{douillard2023diloco, sani2024photon} training approaches enable the utilization of compute resources across geographically distributed data centers, mitigating GPU shortages at any single location~\cite{ye2024openfedllm}. 

While distributed large ML model training presents a promising solution to resource scarcity, it introduces a challenge: communication overhead. For instance, transmitting the 140GB LLaMA-2 model for an all-reduce operation over a 10 Gbps network can add two minutes of communication delay to every training iteration, which significantly increases the total training time.

Prior work has attempted to address this problem in two main ways. The first approach focuses on reducing the volume of data communicated. Systems like Photon~\cite{sani2024photon} and DiLoCo~\cite{douillard2023diloco} leverage techniques like low-rank adaptation and infrequent synchronization to reduce the communication cost of distribution. 
However, these methods typically operate on a \textit{fixed}, pre-configured number of sites and do not adapt dynamically-as we show in \autoref{sec:dynamic_performance_evaluation}, this results in significant opportunity losses.
Further, these systems fail to consider the fact that different amounts of power are available at different sites.

Determining the optimal number of sites to use is not only resource-intensive but also highly dependent on the specific workload. Choosing too few sites negatively affects performance, while choosing too many creates a communication bottleneck. The second approach leverages heuristics for distribution, such as selecting all available resources within a fixed geographical radius to ensure low latency~\cite{strati2024ml, jiang2025thunderserve}. The limitation of such a strategy is that it may ignore sites with more power that lie just outside the radius, and it fails to consider that a site within the radius might lack the power capacity to effectively use its hardware.

To address this problem, we introduce \systemName, a system designed to optimize the power-communication trade-off. 
The limitations above highlight a more fundamental optimization problem. Distributing training across more sites provides access to greater aggregate power, which improves computational speed and reduces time-to-accuracy. However, this distribution also increases communication costs, as model updates must be synchronized across a wider and more dispersed network. At a certain point, the performance penalty from this communication overhead outweighs the benefits of additional computational power. The core challenge is that this optimal balance is not static: it depends on the specific power capacities and network conditions of the available sites at any given time. 
Thus, \systemName's primary goal is to maximize training performance by dynamically selecting the optimal number of sites at runtime. 
\systemName achieves this goal by addressing three fundamental and interconnected questions:
\begin{enumerate}[leftmargin=*, itemsep=0.05cm, topsep=2pt]
\item[\ding{182}] \textbf{\textit{Which sites to choose?}} \systemName ranks all available sites using a power-to-cost heuristic and selects sites that offer the best combination of high power availability (for better performance) and low communication cost.
\item[\ding{183}] \textbf{\textit{How many sites to choose?}} \systemName employs a dynamic greedy heuristic approach and uses the marginal gain in training efficiency (accuracy improvement per unit of time) to identify the optimal number of sites after which the performance penalty from communication overhead negates the benefit of adding more computational power. 
\item[\ding{184}] \textbf{\textit{How to scale the number of sites?}} \systemName implements a greedy bottom-up scaling policy. It begins with a small set of high-power, low-latency sites and incrementally adds more if the marginal gain in training efficiency is positive. This approach avoids the significant initial communication costs of a top-down or fixed / static strategy and ensures that the communication cost per unit of accuracy gain remains low by design.
\end{enumerate}
This paper presents our method for optimizing site selection to minimize training time by balancing the benefits of distributed power with the communication costs of model synchronization. 
To achieve this, \systemName's dynamic optimization is twofold: at runtime, it decides not only \textit{which} sites to select based on their individual efficiency gains, but also \textit{how many} sites to include in the training process while minimizing the network delay. This second decision is important, as the quantity of selected sites determines the total aggregate power and, thus, the speed of computation.

\noindent
\textbf{Contributions.} We summarize our main contributions as follows:
\begin{itemize}[leftmargin=*, itemsep=0.05cm, topsep=2pt]
    \item We formulate and analyze an important trade-off between increased performance from distributed power availability and the communication delays incurred when training large ML models across geographically distributed sites.
    \item We propose \systemName, a novel system that uses a dynamic greedy heuristic to select the optimal number of sites at runtime, maximizing training performance in power-constrained, distributed environments.
    \item We integrate publicly available Google data center power traces~\cite{sakalkar2020data} into our experimental setup. This allows us to model and validate our system against real-world power constraints.
    \item We implement \systemName using the Flower framework~\cite{flower} as an aggregator-based distributed training infrastructure and evaluate it across multiple datasets. Our results demonstrate that \systemName reduces the total training time by \textbf{44\%} against an optimal static policy using uniform power profiles and by \textbf{25\%} against the same static policy using Google power traces. Furthermore, compared to a power-constrained centralized approach, \systemName is more efficient by \textbf{50\%} (with uniform power) and \textbf{27\%} (using the Google trace power). This demonstrates that \systemName efficiently leverages distributed power while mitigating its network overhead.
\end{itemize}

\section{Problem Formulation} 
\label{sec:background}
This section details the power and compute constraints that motivate geo-distributed training, then formalizes the system architecture and its communication costs to define the power-communication trade-off, our system optimizes.

\subsection{Power and resource constraints}
\label{sec:background_power_constraint}
Training state-of-the-art AI models is a highly resource-intensive process that has resulted in primary bottlenecks that restrict performance within a single data center: limited availability of power capacity and of high-performance hardware (e.g., GPUs).

\noindent\textbf{Power capacity model.}
A data center, which we refer to as a site in this paper, is fundamentally constrained by its maximum power supply from a local electric grid. As the computational requirements for AI have grown exponentially~\cite{denning2016exponential} their power demand has begun to exceed the capacity of an individual site.
We model a site's computational throughput---its total processing power---as a direct function of its available power. Let a site $k$ have a maximum power capacity of $P_{cap, k}$ (in watts). For instance, Google reports an average PUE of approximately 1.10, where infrastructure overhead like cooling accounts for only about 10\% of the total power draw~\cite{puegoogle}. Our analysis therefore focuses on the computational power budget, as it is the primary driver of both energy consumption and performance limitations.

The site hosts a number of computational devices, such as GPUs, where each active GPU consumes an average power of $P_{gpu}$. The number of GPUs, $N_{gpu, k}$, that can be concurrently active at site $k$ is therefore constrained by:
$N_{gpu, k} \le P_{cap, k} / P_{gpu}$.
This inequality establishes a limitation on the amount of parallel computation a single site can support. The computational throughput of the site, $C_k$, is directly proportional to the number of active GPUs.

\begin{figure}[t]
    \centering
    \includegraphics[width=\linewidth]{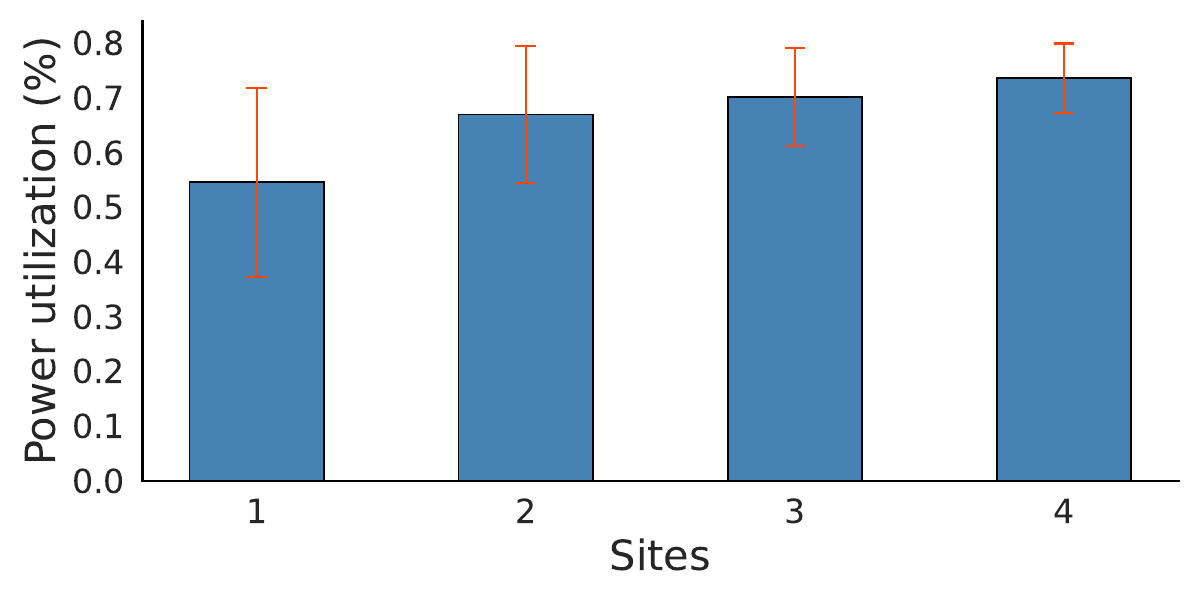}
    \vspace{-0.5cm}
    \caption{\textbf{\emph{The power utilization at different sites in Google's data center power traces. }}}
    \label{fig:pdu_mean_utilization}
    \vspace{-0.7cm}
\end{figure}

\begin{figure}[t]
    \centering
    \includegraphics[width=\linewidth]{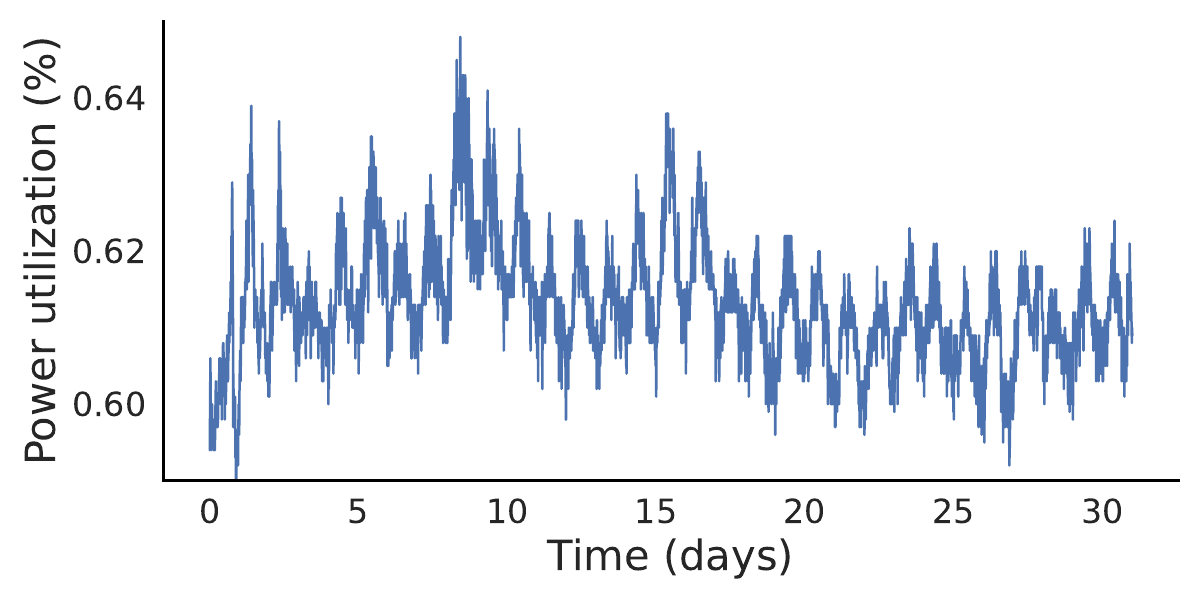}
    \vspace{-0.5cm}
    \caption{\textbf{\emph{Time-series variation of power utilization for a single Power Distribution Unit (PDU) in one of Google's data center.}}}
    \label{fig:production_power_utilization_over_time}
    \vspace{-0.7cm}
\end{figure}

For a realistic model, we leverage publicly available \textbf{Google data center power traces}~\cite{sakalkar2020data}. These traces provide fine-grained, five-minute granularity production power utilization data from large-scale compute cell (equivalent to a site), each containing roughly ten thousand machines. \autoref{fig:pdu_mean_utilization} illustrates the power utilization across various sites. The height of each bar represents the mean power utilization, while the error bars depict the minimum to the maximum measured value for each respective cell. Across the 10 sites, the mean production power utilization varies from 49\% to 74\%, which shows that a site's power utilization is highly dynamic. Additionally, \autoref{fig:production_power_utilization_over_time} shows that the power draw for any given PDU (power distribution unit for each rack of servers) at these sites is highly dynamic over time. This shows that a site may have a large inventory of physically available and idle hardware (e.g., CPUs or GPUs), but if the total power draw is already at its maximum, these resources cannot be fully utilized. This means that a site's effective computational capacity is not determined by its hardware inventory alone, but by its \textit{available power capacity} given by $P_{available, k}(t) = P_{cap, k} - P_{utilized, k}(t)$, at any given time $t$.

\subsection{Geo-Distributed Training with a Central Aggregator}
\label{sec:background_distributed_training}
To leverage computational resources from geographically distributed locations, prior work has explored multiple distributed ML techniques. For instance, federated learning or all-reduced distributed ML synchronization techniques~\cite{douillard2023diloco, sani2024photon, strati2024ml}. However, in this paper, we use a common architecture that involves a central aggregator coordinating a set of distributed worker sites. This model is structurally analogous to Federated Learning but is motivated by resource aggregation rather than data privacy. Additionally we adopt \textbf{data parallelism} in this work as it treats each site as a self-contained "worker" that can contribute to the overall training process. Each site holds a complete replica of the model and trains on its own partition of the data. This makes the system highly modular and scalable. Whereas the primary alternative, model parallelism, involves splitting the model itself across multiple sites, which creates tight dependencies between sites. In a dynamic, geo-distributed environment like ours, adding or removing a site would require a disruptive re-partitioning of the entire model mid-training, which is not feasible. Data parallelism avoids this completely. 

\noindent
\textbf{Distributed training with aggregator.}
The global objective of this technique is to find the optimal model parameters $\theta^*$ that minimize a global loss function $L(\theta)$, defined as a weighted average of the local loss functions $L_k(\theta)$ at each of the $K$ participating sites:
\begin{equation}
    \theta^* = \arg\min_{\theta} L(\theta) \quad \text{where} \quad L(\theta) = \sum_{k=1}^{K} w_k L_k(\theta) 
\end{equation}

Here, $w_k$ is the weight for site $k$, typically proportional to the size of its local dataset $D_k$. 
Training proceeds in synchronous communication rounds, indexed by $t$. In each round, the process begins with a \textit{broadcast} phase, where the aggregator sends the current global model parameters, $\theta^t$, to a selected subset of $S$ sites. Following the broadcast, each site $k$ enters a \textit{local computation} phase, using $\theta^t$ to initialize its local model and performing $E$ epochs of training on its local data $D_k$ to yield an updated local model, $\theta_k^{t+1}$. Subsequently, during the \textit{update communication} step, each site computes its local update, $\Delta_k^{t+1} = \theta_k^{t+1} - \theta^t$, and transmits it back to the aggregator. Finally, the round concludes with the \textit{aggregation} phase, where after receiving updates from all $S$ sites, the aggregator computes the new global parameters using the formula:
\begin{equation}
    \theta^{t+1} = \theta^t + \sum_{k=1}^{S} w_k \Delta_k^{t+1}
\end{equation}

\noindent
\textbf{Communication delays.}
While geo-distributed training presents a solution to resource limitation, it introduces the significant challenge of communication overhead. In a synchronous distributed system, the total time for a single training round ($T_{\text{round}}$) is the sum of the computation time ($T_{\text{compute}}$) and the communication time ($T_{\text{comm}}$). The communication time is dominated by the need to synchronize model parameters or updates across wide-area networks (WANs). 
For a model update of size ($D_m$) being sent from a site $k$ to a central aggregator, the communication time is at least:

\vspace{-0.2cm}
\begin{equation}
\label{eq:communication_per_site_delay}
    T_{k, \text{comm}} \ge \frac{2d_k}{c} + \frac{D_m}{B_k}
\end{equation}

where $d_k$ is the physical distance of site $k$ to the aggregator, $c$ is the speed of light in the propagation material, and $B_k$ is the available bandwidth. In a synchronous round over a set of sites $S$, the system is bottlenecked by the slowest site (the "straggler"), making the round's communication time $T_{\text{comm}}(S) = \max_{k \in S} \{T_{k, \text{comm}}\}$. 
The total communication time for the round is then determined by the maximum of these individual times over the set of selected sites $S$:

\vspace{-0.2cm}
\begin{equation}
    T_{\text{comm}}(S) = \max_{k \in S} \left\{ \frac{2d_k}{c} + \frac{D_m}{B_k} \right\}
\end{equation}

Given that inter-continental latencies can be over 100ms and model updates can be gigabytes in size, this overhead can easily nullify the performance gains from more power availability, making the site selection optimization an important research problem. 

\subsection{The Performance Trade-off}
\label{sec:background_performance_tradeoff}
The overall goal of our system is to reach a target accuracy in the least amount of time (i.e., minimize the \textbf{time-to-accuracy}) given the power constraints at a single site, greater aggregate power across multiple sites, and the communication overheads. 
This can also be framed as maximizing the training efficiency per unit of time. 
We define training efficiency, $\eta$, as the accuracy gain per second:

\begin{equation}
\label{eq:efficiency}
    \eta(S) = \frac{\mathbb{E}[\text{Accuracy Gain per Round}]}{T_{\text{round}}(S)}
\end{equation}

where the total round time for a set of sites $S$ is a function of both computation and communication. The computation time depends on the effective throughput of the selected sites, which is constrained by both hardware availability and real-time power capacity.
Following the model established from the Google power traces, the effective number of GPUs at a site $k$ at time $t$, denoted $N_{eff, k}(t)$, is the minimum of the physically available GPUs, $N_{avail, k}$, and the number of GPUs that can be supported by the site's available power at that moment, $P_{avail, k}(t)$. We formalize this as:

\vspace{-0.2cm}
\begin{equation}
    N_{eff, k}(t) = \min\left(N_{avail, k}, \left\lfloor \frac{P_{avail, k}(t)}{P_{gpu}} \right\rfloor\right)
\end{equation}

Therefore, the site's effective throughput at time $t$ is directly proportional to this number: $ C_k(t) = N_{eff, k}(t) \cdot C_{single\_gpu} $ where $C_{single\_gpu}$ is the processing power of the single GPU. This model  captures the scenario where a site's computational output is limited by its power budget, even if it has a surplus of idle hardware.
Thus the total round time $ T_{\text{round}}(S) = T_{\text{compute}}(S) + T_{\text{comm}}(S)$ is:
\begin{equation}
\label{eq:round_time_equation}
T_{\text{round}}(S) \propto \frac{1}{\sum_{k \in S} C_k(t)} + \max_{k \in S} \left\{ \frac{2d_k}{c} + \frac{D_m}{B_k} \right\}
\end{equation}
This formulation makes the trade-off explicit: as we increase the number of sites $|S|$, we aggregate their independent power budgets. This increases the total effective computational throughput $\sum C_k(t)$ by allowing more of the available hardware to be utilized, which decreases the computation time, $T_{\text{compute}}(S)$, and in turn increases training efficiency. At the same time, as $|S|$ increases, the maximum communication time, is non-decreasing and likely to increase as more distant or lower-bandwidth network sites are included. This increases the total round time, decreasing training efficiency.

Due to these opposing effects, there exists an optimal subset of sites, $S^*$, that maximizes the training efficiency $\eta(S^*)$. Choosing too few sites leaves valuable computational power untapped, prolonging training. Choosing too many sites results in the system being bottlenecked by communication delays, again prolonging training. The challenge, therefore, is to dynamically identify this optimal set $S^*$ at runtime. \systemName optimizes this tradeoff to maximize performance gains under power constrained environments.

\section{\systemName Design}
\label{sec:design}

\begin{figure}[t]
    \centering
    \includegraphics[width=\linewidth]{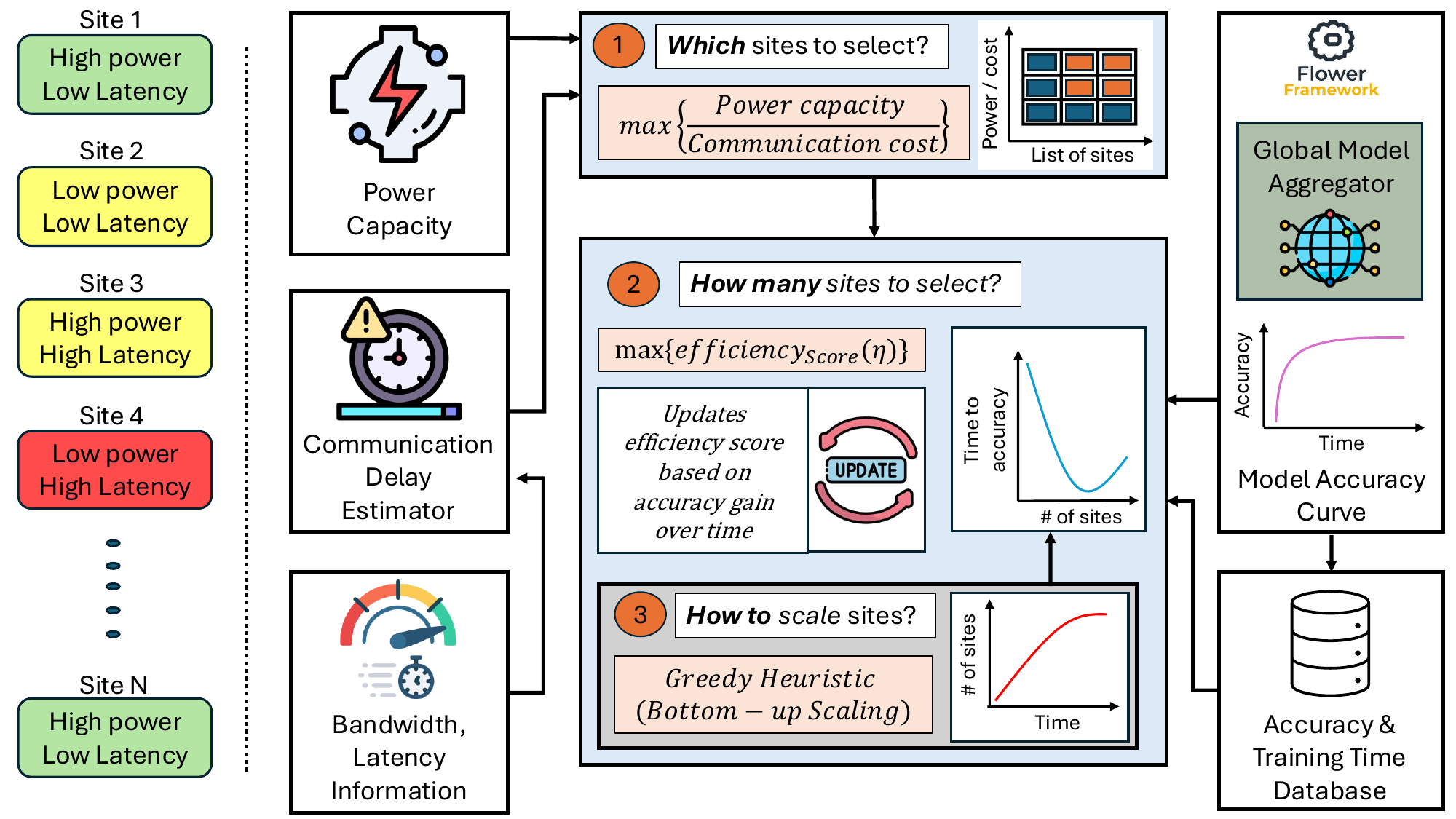}
    \vspace{-0.4cm}
    \caption{\textbf{\emph{\systemName's system design and its components.}}}
    \label{fig:system_design_figure}
    \vspace{-0.7cm}
\end{figure}

\autoref{fig:system_design_figure} provides an overview of \systemName's design, which solves the power-communication trade-off by intelligently and dynamically selecting the optimal subset of available sites at runtime to maximize training performance. \systemName is designed to integrate with and expand upon existing distributed learning frameworks. 

Existing frameworks~\cite{tff2019, ryffel2018generic} already handle fundamental functions such as communication, site configuration, and model aggregation. However, these frameworks typically lack native support for profiling power availability and network delays (bandwidth and latencies) of distributed sites, which we enable by implementing the necessary auxiliary modules. As detailed in~\autoref{sec:implementation}, we chose the Flower framework~\cite{flower} for our implementation due to its flexible API, which allowed us to integrate our dynamic site-selection policy. While we use Flower, our design is framework-agnostic and can be applied to any system that allows for the configuration of selected sites in each synchronous training round.

We next detail \systemName's modules and describe how they enable it to make three core decisions that optimize the power-communication trade-off:

\begin{enumerate}[leftmargin=*, itemsep=0.05cm, topsep=2pt]
    \item Which sites to select to maximize performance by exploiting high power availability and low communication cost? (\autoref{sec:design_which})

    \item How many sites to choose to identify the optimal number of sites, after which the performance penalty from communication overhead negates the benefit of more power availability?  (\autoref{sec:design_how_many})
    
    \item How to scale the number of sites to ensure that the communication cost per unit of accuracy gain remains low? (\autoref{sec:dynamic_policy})
\end{enumerate}

In presenting these modules, we highlight the design decisions that enable \systemName to dynamically optimize heterogeneous power-communication trade-offs.

\subsection{Site selection for performance gains}
\label{sec:design_which}
\systemName’s site selection strategy aims to rank all available sites using a power-to-cost heuristic and then selects sites that offer the best combination of high power availability (for better performance) and low communication costs. This strategy draws inspiration from two simpler heuristics: the lowest network latency heuristic and the highest power heuristic, which serve as baselines.

A site $k \in S$ is characterized by its effective power availability, $P_{avail, k}(t)$, and its communication delay to the aggregator, $T_{k, \text{comm}}$. A selection strategy can prioritize one or both of these metrics:

\noindent
\textbf{Lowest Latency.}
This strategy prioritizes minimizing communication overhead, $T_{k, \text{comm}}$ as given in \autoref{eq:communication_per_site_delay}, by creating a ranked list of candidates based on their individual communication delays. The final output of this strategy is an ordered list of all available sites, with those exhibiting the best connectivity (lowest latency) positioned at the top. It is designed for scenarios where the overall training time is dominated by communication bottlenecks, such as in geographically dispersed or bandwidth-constrained environments where fast synchronization is important. However, its primary weakness is that it does not explicitly consider computational power, $P_{avail, k}$ and as we established in \autoref{sec:background_power_constraint}, power availability varies significantly across sites. Consequently, this policy will perform poorly if the sites with the lowest communication cost are also the ones with the lowest power capacity, as the training would then be limited by slow computation.

\noindent
\textbf{Highest Power.}
This strategy prioritizes maximizing power availability by ranking all available sites $k \in \mathcal{S}$ solely based on their effective computational power, $P_{avail, k}$. The resulting output is an ordered list of the sites with most power available. This approach is designed for scenarios where local computation is the primary bottleneck, such as when training large scale models. Because this strategy does not explicitly consider network latency, it is most suitable when all sites have high-bandwidth, low-latency connections, making communication time a negligible factor in the overall training duration. However, as discussed in~\autoref{sec:background_performance_tradeoff} the communication delays can be significant in a distributed training environment, thus this policy is likely to select a powerful data center that is geographically distant, which then becomes a high-latency ``straggler'' in every synchronous training round. This forces all other faster sites to wait and delays the training time. 

\noindent
\textbf{\systemName's selection strategy.}
Given that strategies focusing on a single metric are prone to creating computational or communication bottlenecks, \systemName's selection strategy provides a balanced approach that simultaneously considers both power availability and the communication delays. However, we recognize that the relative importance of computational power versus communication cost is not static; it can change depending on the size of the model update being transmitted, given that bandwidth remains the same. To account for this, our strategy uses an adaptive, weighted scoring mechanism. 

The idea is to calculate a weighted power-to-cost score for each site $k \in \mathcal{S}$. This score is designed to be flexible, adjusting its priorities based on the characteristics of the training workload. The score is calculated as:
\begin{equation}
    \text{Power-to-cost score}(s_i) = \frac{P_{avail, k} \cdot {w_p}}{T_{k, \text{comm}} \cdot {w_c}}
\end{equation}

where $w_p$ and $w_c$ are dynamic weights that control the emphasis on power versus communication cost. These weights are adjusted based on the model update size, $D_m$, to adapt the selection heuristic to the workload. For small update sizes, where overall training time is more sensitive to local computation speed, the system prioritizes power by setting $w_p > w_c$. Conversely, for large updates sizes, where network latency becomes the dominant bottleneck, it prioritizes low communication cost by setting $w_c > w_p$. This allows the ranking to effectively identify the most valuable sites for different training scenarios.

\subsection{Site provisioning and power-cost trade-off}
\label{sec:design_how_many}
Once a site selection strategy generates a ranked list of candidates with the best power-to-cost score, the next design question is to determine the optimal number of sites to select during the training to balance the benefit of more computational power against the cost of higher communication delays. In this section, we explore different policies for making this decision, which serve as baselines for \systemName's dynamic policy, which we discuss in \autoref{sec:dynamic_policy}.

\noindent
\textbf{Select all sites.}
This strategy represents a simple baseline where the system utilizes every available site for training ($S' = \mathcal{S}$), assuming that maximizing parallel computation also yields maximum performance. However, this approach fails to consider the impact of communication overhead in a synchronous environment.

As shown in our round time model in \autoref{eq:round_time_equation}, while selecting all sites maximizes the aggregate power ($P_{avail}$) and minimizes the computational time $T_{\text{compute}}(S)$, it also maximizes the communication time $T_{\text{comm}}(S)$, which is dictated by the site with the highest latency in the entire set of  selected sites ($\max_{k \in \mathcal{S}} T_{\text{comm,k}}(S)$). In any large-scale, geographically distributed system, this value creates a synchronization bottleneck, forcing all high-power, low-latency sites to idly wait. As a result, the training time for each round becomes dominated by communication, which reduces overall performance.

This strategy serves as a useful baseline, demonstrating that blindly maximizing computational resources is an ineffective approach and motivates a more intelligent selection mechanism.

\noindent
\textbf{Static selection.}
This policy is an offline, analytical approach designed to empirically find the theoretical optimal \textit{fixed} number of sites for a given workload. This policy directly explores the power-communication trade-off by running a series of exhaustive, end-to-end training experiments with the primary objective of finding the number of sites, $k$, that minimizes the total time required to reach a target model accuracy ($T_{\text{target\_acc}}$). This time-to-accuracy is a product of the number of synchronization rounds needed ($R_{\text{acc}}$) and the average time per round ($T_{\text{round}}$), given by: $T_{\text{target\_acc}}(S_k) = R_{\text{acc}}(S_k) \cdot T_{\text{round}}(S_k)$

An optimal number of sites, $k^*$, exists due to the competing effects on this time-to-accuracy as $k$ increases. As more sites are added to the training group ($S_k$), the aggregate computational power ($\sum P_{avail,k}$) increases, which reduces the computation time per round. In contrast, including more sites (which are selected from progressively farther away) increases the likelihood of high-latency sites, which increases the communication time per round ($\max (T_{\text{comm}}(S))$). This creates a distinct trade-off where the initial benefits of adding more power are eventually outweighed by the communication costs, ensuring that a minimum value for $T_{\text{target\_acc}}$ exists at some optimal $k^*$.

This creates a distinct trade-off. Initially, for small $k$, adding sites dramatically reduces training time because the computation gains are significant and outweigh the minor increases in latency. However, after a certain point, there are diminishing returns. The added sites are farther away, contributing less marginal computational power while significantly increasing the communication bottleneck. This causes the round time, and consequently the total training time, to increase. This behavior ensures that a minimum value for $T_{\text{total\_acc}}$ exists at some optimal $k^*$.

This approach for finding the optimum is straightforward but resource-intensive. For a chosen site selection strategy, we first generate a ranked list of all available sites. We then execute the full training process for various group sizes, $k \in \{k_1, k_2, \dots, S\}$ sites. For each complete run, we measure the final time-to-accuracy. The optimal fixed number of sites, $k^*$, is then identified as the value of $k$ that resulted in the lowest total training time:
\begin{equation}
k^* = \underset{k}{\arg\min} \ T_{\text{target\_acc}}(S_k)
\end{equation}

The primary drawback of this policy is that it is an \textit{offline, post-hoc analysis}, making it too resource-intensive. However, it provides an important baseline against which we evaluate \systemName's dynamic policy, as the resulting $k^*$ represents the best performance achievable with any \textit{fixed set of sites}.

\begin{table}[t]
\centering
\caption{Variables in \systemName's dynamic policy}
\label{tab:variables}
\footnotesize
\vspace{-0.3cm}
\begin{tabular}{ll}
\toprule
\textbf{Variable} & \textbf{Description} \\
\midrule
$k$ & A candidate number of sites to evaluate \\
$k_{\text{current}}$ & The number of sites used in the current round \\
$\mathcal{K}$ & The set of candidate counts for the next phase \\
$\alpha_r$ & Model accuracy at the end of round $r$ \\
$\Delta\alpha_k$ & Projected accuracy gain for using $k$ sites \\
$\tau_k$ & Estimated round time for using $k$ sites \\
$\eta_k$ & Training efficiency score for using $k$ sites \\
$\alpha_{\text{target}}$ & Target accuracy threshold to stop training \\
$p$ & Patience: number of rounds to wait before reducing sites \\
\bottomrule
\end{tabular}
\vspace{-0.3cm}
\end{table}

\subsection{\systemName's Dynamic Optimization Policy}
\label{sec:dynamic_policy}

The dynamic optimization policy is the core of our proposed system, \systemName. Unlike the fixed Static Policy, this approach determines the optimal number of training sites at runtime, continuously adapting to the real-time power and network conditions of the distributed environment. Our goal is to actively manage the power-communication trade-off during training to minimize the overall time-to-accuracy. 

\noindent
\textbf{Greedy heuristic for runtime optimization}
As detailed in~\autoref{alg:dynamic_selection}, the training process begins by initializing the active site count to a small number, $k_{\text{current}} = \min(k_{current}, S)$, where $S$ is the total number of available sites. These sites are the best-ranked sites, i.e., those with the highest Power-to-Cost Score as discussed in~\autoref{sec:design_which}. After the first training round is complete, the system must decide how to scale the number of participants for the next round. To address this, our system adopts a \textbf{bottom-up, greedy incremental addition strategy}. This approach starts with a small, core group of the best-ranked sites, i.e., those with the highest Power-to-Cost Score. It then greedily adds more sites in subsequent rounds. This strategy is more efficient than a top-down, subtractive method that begins with all available sites as the initial training rounds would be slow due to the communication bottleneck from including all high-latency sites.

The system then evaluates a set of potential site counts, $\mathcal{K}$, within a window $k_{\text{current}} \pm n$ around its current count $k_{current}$. For each candidate count $k \in \mathcal{K}$, the system projects the expected accuracy gain, $\Delta\alpha_k$, using a model that incorporates recent accuracy trends ($\alpha_r - \alpha_{r-1}$), a normalization factor $k/k_{\text{current}}$ for the number of workers, and a logarithmic term to account for diminishing returns. 

Concurrently, the system estimates the round time $\tau_k$ as the sum of two components: the computation time, which is inversely proportional to the mean power capacity (or computational capability) of the selected sites, and the communication time, which is determined by the maximum network latency among the selected sites. Specifically, $\tau_k = \frac{T_{\text{comp}}}{\text{mean}(P_{avail(1:k)})} + \max(T_{comm(1:k)})$, where $P_{avail}$ represents the power capacity at a single location and $T_{comm}$ represents the network latency for the top $k$ sites.

At each synchronization round, the optimal site count $k'$ is then selected by maximizing the accuracy-efficiency gain $\eta_k = \frac{\Delta\alpha_k}{\tau_k}$ across all candidates $k \in \mathcal{K}$. This ratio effectively balances accuracy improvements against time costs, ensuring the system selects the site count $k'$ that delivers the best accuracy-per-time-unit performance until the system reaches the target accuracy.

\begin{algorithm}[t]
\caption{Greedy Heuristic Dynamic Site Selection}
\label{alg:dynamic_selection}
\DontPrintSemicolon
\SetKwInOut{Input}{Input}
\SetKwInOut{Output}{Output}
\Input{$S$ sites ranked by power-cost-score $s_i = \frac{P_{avail}}{T_{comm}}$ ($P_{avail}$: available power, $T_{comm}$: network latency)}
\Output{Trained model parameters after reaching $\alpha_{\text{target}}$}

\textbf{Initialization:}\;
$k_{\text{current}} \gets \min(\text{initial\_sites}, S)$\; 
$r \gets 0$, $\text{last\_improvement} \gets 0$\; 

\While{$r \leq r_{\text{max}} \text{ and } \alpha_r < \alpha_{\text{target}}$}{
    \textbf{Training Round:}\;
    Select top $k_{\text{current}}$ sites from ranked list\;
    $\alpha_r, \tau_r \gets \text{TrainModel}(k_{\text{current}})$\; 
    
    \If{$r - \text{last\_improvement} > p$}{
        \textbf{break}\; \tcp{Stop if accuracy has plateaued}
    }
    
    \If{$r \mod \text{adjustment\_interval} = 0$}{
        $\mathcal{K} \gets \text{GetCandidateCounts}(k_{\text{current}}, K)$\; \tcp{e.g., $k_{\text{current}} \pm n_1, \pm n_2$}
        
        \For{\textbf{each} $k \in \mathcal{K}$}{
            $\tau_k \gets \frac{T_{\text{comp}}}{\text{mean}(P_{avail}(1:w))} + \max(T_{comp(1:k)})$\;
            $\Delta\alpha_k \gets \text{ProjectAccuracyGain}(k, k_{\text{current}}, \alpha_r, \alpha_{r-1})$\; \tcp{See text for formula}
            $\eta_k \gets \frac{\Delta\alpha_k}{\tau_k}$\; 
        }
        
        $k^* \gets \underset{k \in \mathcal{K}}{\operatorname{argmax}} \,\eta_k$\; 
        \If{$k^* < k_{\text{current}} \text{ and } r - \text{last\_improvement} < p$}{
            $k^* \gets k_{\text{current}}$\; 
            \tcp{Avoid premature reduction}
        }
        $k_{\text{current}} \gets k^*$\;
    }
    
    \If{$\alpha_r > \alpha_{r-1}$}{
        $\text{last\_improvement} \gets r$\;
    }
    $r \gets r + 1$\;
}
\end{algorithm}

\noindent\textbf{Advantage in time-to-accuracy.}
The primary advantage of our dynamic policy over the static policy lies in its better efficiency over the entire training life-cycle. A static policy using a large, fixed number of sites, $k^*$, is often bottlenecked by the high communication cost of its slowest site in every single round. The total training time is the sum of these consistently slow rounds: $T_{\text{static}} = \sum_{r=1}^{R_{\text{static}}} T_{\text{round}}(S_{k^*})$.

In contrast, our dynamic policy begins with a small, highly efficient group of sites, $k_0$, where the maximum latency is minimal, making the initial training rounds faster, with relative accuracy gain compared to the $k^*$. The policy then intelligently scales the number of sites, $k_r$, based on its runtime efficiency heuristic. The total time is thus a sum of varying round times, many of which are much faster than the static policy's rounds:
$T_{\text{dynamic}} = \sum_{r=1}^{R_{\text{dynamic}}} T_{\text{round}}(S'_{k_r})$.
By starting small and scaling up when efficient, the dynamic policy reduces training time in the early and middle stages of training.

\section{Implementation}
\label{sec:implementation}
We implement \systemName using Flower~\cite{flower}, a widely-used open-source framework for federated and distributed learning, which enables us to focus on \systemName's core logic rather than its underlying communication protocols. Our implementation follows a centralized architecture composed of a single aggregator, which hosts the core system logic, and multiple distributed sites, each representing a training site. The aggregator organizes the entire training process, including profiling the client sites, executing the dynamic site selection algorithm, and aggregating model updates. Sites receive the global model, train locally, and report model updates and status to the aggregator. \systemName supports QLoRA~\cite{dettmers2023qlora} or LoRA~\cite{hu2022lora} optimizations for training large language models, reducing network overhead by reducing update parameter sizes. 

\systemName is composed of several key modules. A \textit{Site Profiler module}, running on the aggregator, gathers and maintains the information about each available site. For our experiments, we model two key characteristics for each site $k$, using both controlled configurations and real-world data traces. For controlled experiments, the computational capacity, $C_k$, is a pre-configured parameter set at the start to simulate sites with different maximum power capacities. To emulate real-world power capacities, we also leverage Google's publicly available power traces. From these traces, our Site Profiler computes the spare power capacity available for computation at any given time. This allows our system to model a realistic scenario where a site may have sufficient idle compute resources, but its performance is ultimately limited by the lack of available power. The profiler is designed to handle this discrepancy, ensuring that site selection is based on actual power availability, not just hardware inventory. In both scenarios, the network characteristics, $L_k$ and $B_k$, are managed using network emulation tools to control the latency and bandwidth between each client site and the aggregator. The Site Profiler periodically measures these network values to dynamically update the communication cost model used by the selection logic.

The \textit{Site selection module}, which is the core of our contribution, implements our dynamic greedy heuristic. At the start of each selection phase, it ranks all available sites based on a performance-to-cost heuristic that prioritizes high computational throughput and low communication cost. The module then builds the optimal set of sites incrementally. Starting with the top-ranked site, it iteratively considers adding the next-best site from the list. A site is only added if the system estimates that its inclusion will improve the overall training efficiency compared to the previously measured performance. This process continues until adding more sites offers no further benefit, yielding the optimal set for the next training phase.
Finally, for training execution and monitoring, the aggregator uses standard mechanisms to dispatch tasks to the clients selected by the Site Selection Module. We instrumented the aggregator to measure the wall-clock time for each training round and to evaluate the global model's accuracy on a held-out validation set after each round. This allows the system to compute the realized training efficiency, $\eta_{measured}$ (~\autoref{eq:efficiency}), which, as discussed in \autoref{sec:design}, serves as the feedback signal for the selection module in the subsequent decision-making cycle.

\section{Evaluation}
\label{sec:evaluation}
Below, we present our evaluation methodology (\autoref{sec:eval_setup}), evaluate \systemName's design components (\autoref{sec:power_availability_heterogeneous} and \autoref{sec:dynamic_performance_evaluation}), and analyze \systemName's robustness (\autoref{sec:generalization}).

\subsection{Evaluation Methodology}
\label{sec:eval_setup}

We present our evaluation methodology, covering (1) compute resources, (2) training datasets and models, (3) baseline policies, (4) evaluation metrics and (5) power and communication profiling.

\begin{table}[h]
    \vspace{-0.2cm}
    \centering
    \small
    \caption{Statistics for datasets used in the evaluation. }
    \vspace{-0.2cm}
    \label{tab:dataset-stats}
    \begin{tabular}{|c|c|c|c|c|}
    \hline
    \textbf{Dataset} & \textbf{Task} & \textbf{Model} & \textbf{Classes} & \textbf{Samples} \\ \hline \hline
    EMNIST          & Image Class.    & CNN            & 62               & 814,255          \\ \hline
    Shakespeare     & Next-Char Pred. & LSTM           & 80               & 422,615          \\ \hline
    Google Speech   & Keyword Spot.   & KWT-1          & 35               & 105,829          \\ \hline
    \end{tabular}
\end{table}

\noindent\textbf{Experimental setup.} 
We evaluate \systemName on a local cluster of 40 \texttt{NVIDIA GTX 1080Ti} GPUs. We emulate a distributed machine learning environment consisting of 100 unique sites, 
each with distinct training and communication delays. The underlying communication protocols and multi-node orchestration are provided and managed by the Flower framework~\cite{flower}.

\noindent\textbf{Simulated time and resource emulation.}
Given that the number of emulated sites exceeds our available physical GPUs, we employ a simulation strategy to model performance under perfect parallelism. In our setup, a single physical GPU may be responsible for executing the training tasks for multiple virtual sites. These tasks run sequentially on the hardware; however, we calculate a \textit{simulated wall-clock time} for each round to accurately reflect how the system would perform if each site had its own dedicated resources running in parallel. This environment also allows us to simulate and evaluate different parameter update sizes which affect the communication time. The total time for a site $k$ is the sum of its computation time and communication time. The communication time is described in~\autoref{eq:communication_per_site_delay} whereas the computation time is determined by scaling the emulated time by the site's emulated power availability ($P_{avail, k}$). This method allows us to evaluate our policy's ability to handle resource heterogeneity at a scale that would otherwise be infeasible, while accurately modeling the performance bottlenecks that arise in a real parallel execution environment.

\noindent\textbf{Datasets and models.}
We evaluate our approach on three datasets representing distinct tasks and complexities. For the EMNIST image recognition dataset~\cite{emnist-dataset}, we train a simple Convolutional Neural Network (CNN) from scratch with a learning rate of 0.01. For next-character prediction on the Shakespeare text corpus~\cite{caldas2018leaf}, we use a two-layer LSTM~\cite{li2020federated} with a learning rate of 0.8. Finally, for speech classification on the Google Speech Commands dataset~\cite{warden2018speech}, we train a KWT-1 model~\cite{berg2021keyword} with a learning rate of 0.001. The batch size was 20 for EMNIST and Shakespeare, and 16 for Google Speech. The characteristics of each dataset are summarized in \autoref{tab:dataset-stats}. 

\begin{figure}[t]
    \centering
    \includegraphics[width=0.9\linewidth]{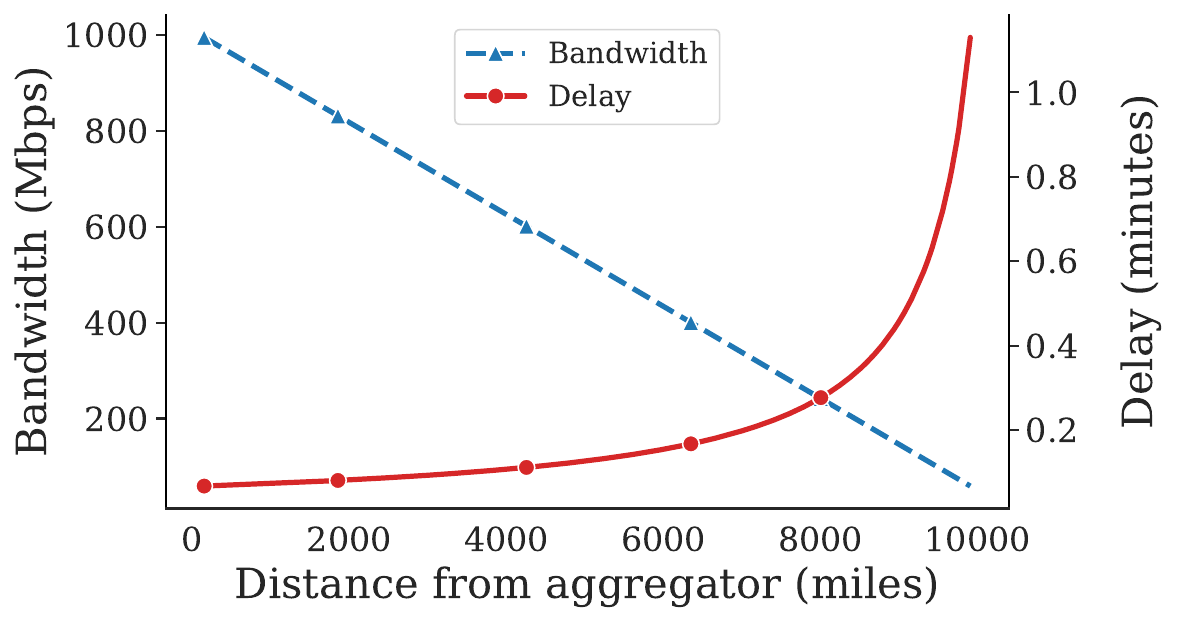}
    \vspace{-0.3cm}
    \caption{\textbf{\emph{Bandwidth decreases where as communication delay increases with increasing geographical distances.}}}
    \label{fig:bandwidth_delay_distance}
    \vspace{-0.2cm}
\end{figure}

\begin{figure}[t]
    \centering
    \includegraphics[width=0.9\linewidth]{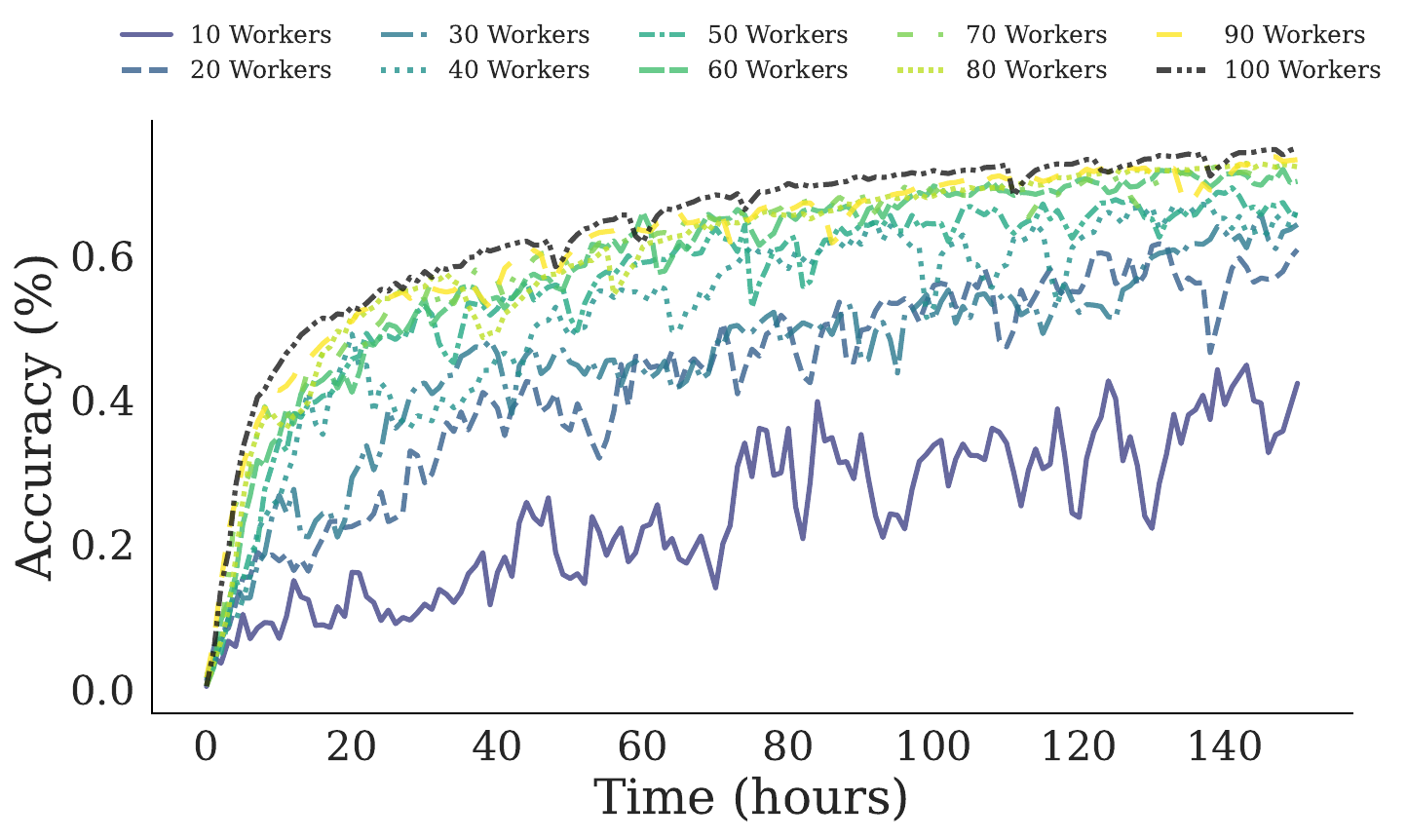}
    \vspace{-0.3cm}
    \caption{\textbf{\emph{Using more sites improves convergence accuracy due to the greater aggregate computational power available, assuming the same power and network delay across all sites. }}}
    \label{fig:accuracy_vs_time}
    \vspace{-0.5cm}
\end{figure}

\begin{figure*}[t]
    \centering
    \begin{subfigure}[b]{0.33\linewidth}
        \includegraphics[width=\linewidth]{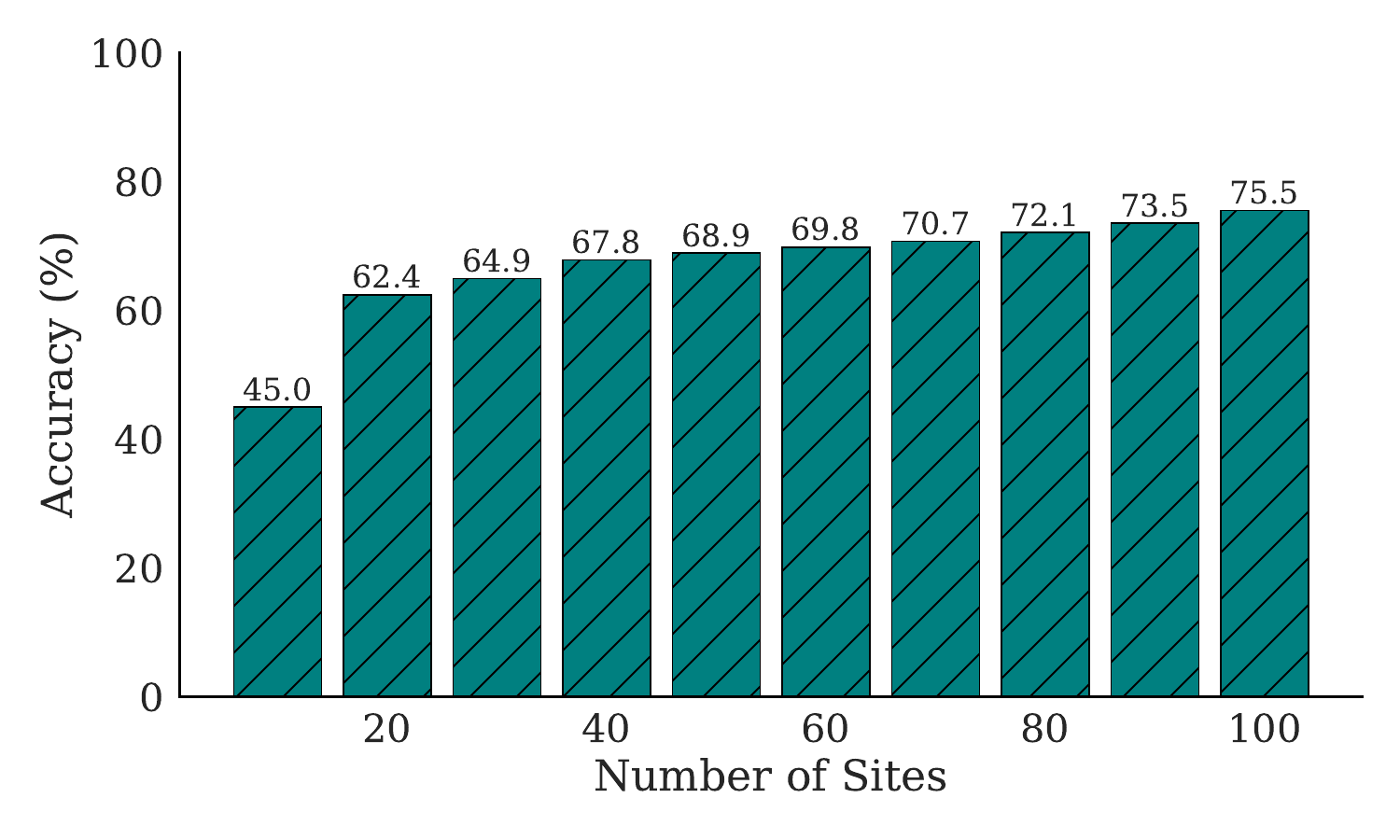}
        \vspace{-0.3cm}
        \caption{Final Accuracy (\%)}
        \label{fig:bar_accuracy}
        \vspace{-0.3cm}
    \end{subfigure}
    \hfill 
    \begin{subfigure}[b]{0.33\linewidth}
        \includegraphics[width=\linewidth]{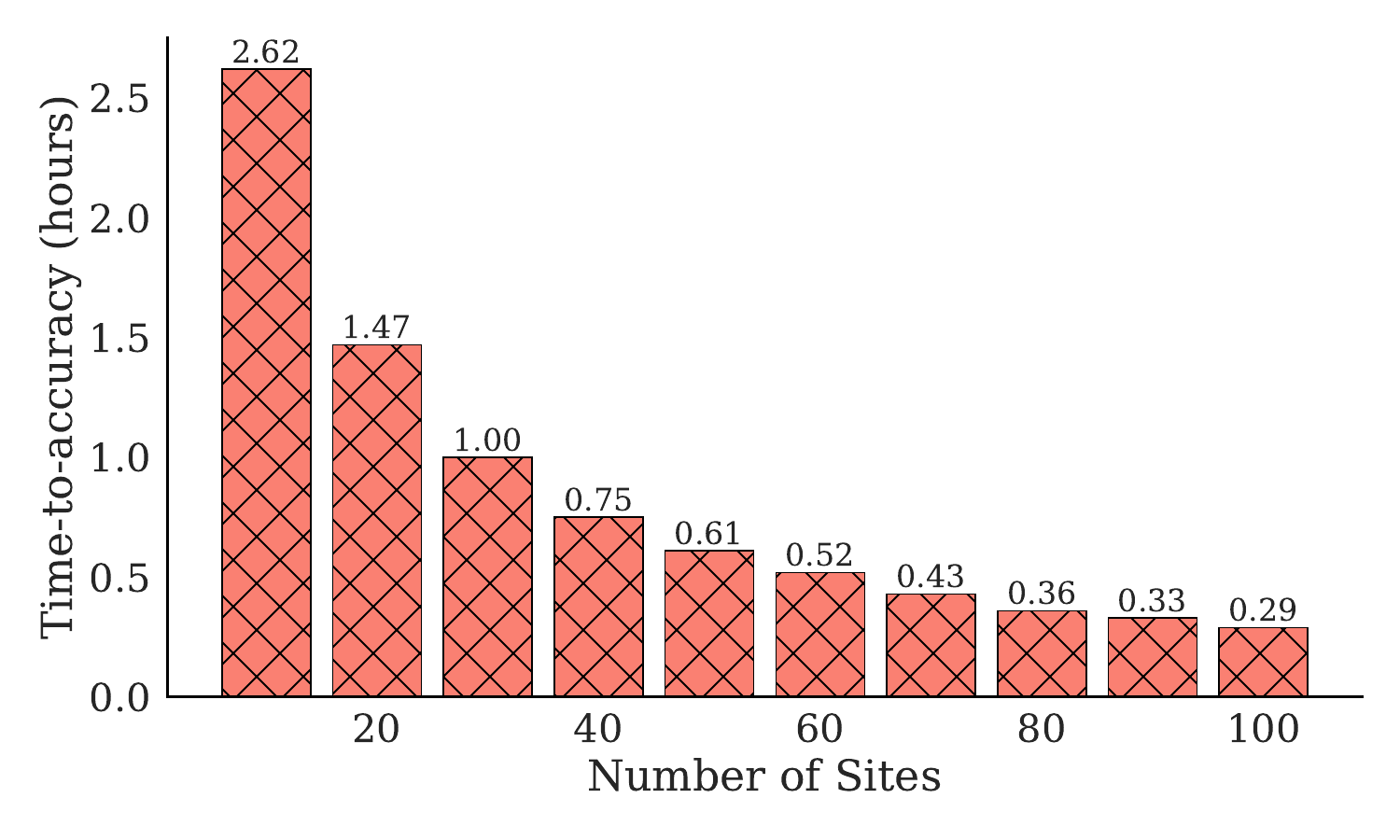}
        \vspace{-0.3cm}
        \caption{Time-to-accuracy (hours)}
        \label{fig:bar_time_to_accuracy}
        \vspace{-0.3cm}
    \end{subfigure}
    \hfill 
    \begin{subfigure}[b]{0.33\linewidth}
        \includegraphics[width=\linewidth]{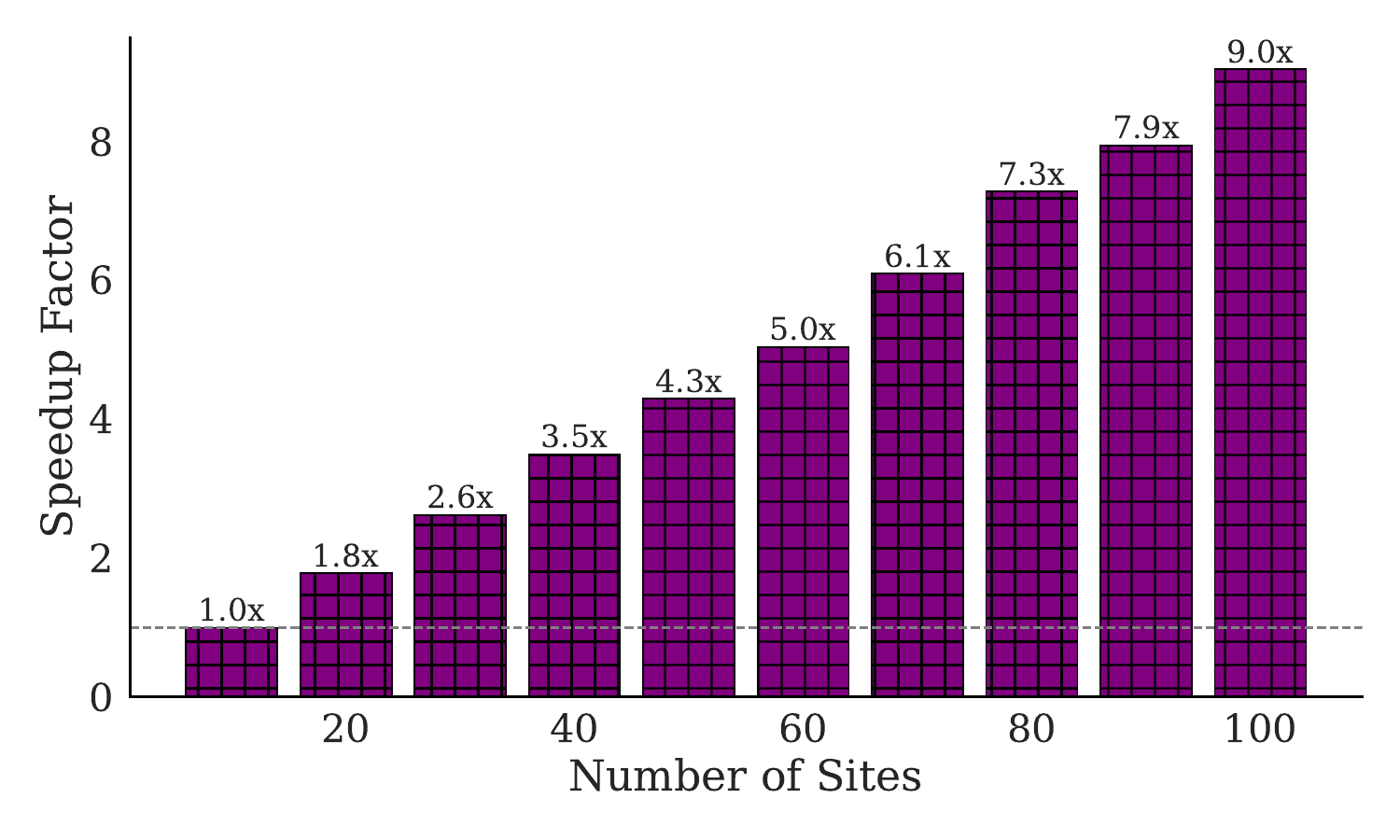}
        \vspace{-0.3cm}
        \caption{Time Speedup Factor}
        \label{fig:bar_speedup}
        \vspace{-0.3cm}
    \end{subfigure}
    
    \caption{\textbf{\emph{Performance metrics - (a) convergence accuracy, (b) time-to-accuracy and (c) speedup factor, improve with more sites, due to greater aggregate computational power available.}}}
    \label{fig:combined_performance}
    \vspace{-0.3cm}
\end{figure*}

\noindent\textbf{Baseline policies.}
Below, we present the baseline approaches that we compare \systemName with.

\begin{itemize}[leftmargin=*, topsep=0.1cm, itemsep=0.05cm]
    \item \textbf{Random selection.} As a baseline for comparison, we use \textit{Random Selection}, a common approach in distributed learning~\cite{fedavg-mlsys, wu2022sustainable}. In this method, a random subset of available sites is chosen to participate in each training round. This randomness is a primary weakness in a performance-oriented system. Random selection is agnostic to the resource profiles of the sites, meaning it might select a site with low computational throughput due to power limitations or a site with high network latency, impacting the performance negatively in both cases.

    \item \textbf{Power constrained centralized training.}
    In this scenario, we emulate the training process occurring at a single site. The primary advantage of this approach is no network communication overhead, as all computation happens at one location. The drawback, however, is that performance is strictly limited by the power capacity of that single site. Even if more hardware is physically available, it cannot be utilized once the site's power cap is reached, thus capping the maximum computational throughput~\cite{patterson2021carbon, avelar2023ai}. This baseline is important as it quantifies the performance capacity for any non-distributed approach and highlights the  motivation for geo-distribution: accessing more aggregate power, while incurring communication costs.
    
    \item \textbf{Fixed radius.} 
    Our second baseline is \textit{Fixed Radius Selection}, which selects all available sites within a predefined geographical distance. The primary advantage of this strategy is low communication latency due to the proximity of sites, leading to faster synchronization rounds. Its drawback, however, is being constrained by the power resources available within that local area, as it may ignore more sites with more power that are farther away. This can result in a system that is bottlenecked by computation despite its efficient communication. This approach is motivated by prior work~\cite{strati2024ml} that suggests keeping data parallelism within a single region to avoid high communication costs. However, that work does not analyze the potential performance gains from accessing greater aggregate power at the cost of higher latency, making this an essential baseline for our study.

    \item \textbf{Static Selection.}
    This baseline represents a static approach where a fixed number of sites, $k$, is selected at the start of training and used for every subsequent round. This fixed number of site selection is observed in prior work~\cite{douillard2023diloco, sani2024photon} as well. The drawback of the fixed-static policy is that determining the optimal number of sites to select is not only resource-intensive but also highly dependent on the specific workload. Choosing too few sites negatively affects the performance, while choosing too many creates a communication bottleneck. However, it provides an important baseline against which we evaluate \systemName's dynamic policy, as the set of selected sites with this policy represents the best performance for a \textit{fixed set of sites}. 
\end{itemize}

\noindent\textbf{Power profiling.}
We model power constraints using two different methods. For controlled, repeatable experiments, we assign each site a static power availability value, drawn from a uniform random distribution between 0.1 and 1.0. This value represents the fraction of a site's maximum potential computational throughput that is currently available. We then simulate the impact of this power constraint on the local training time. For instance, if a site's ideal training time (at 100\% power) is $T_{ideal}$ and its power availability is $P_{avail}$, its effective training time becomes $T_{ideal} / P_{avail}$. This means a site with 50\% power availability ($P_{avail}=0.5$) would take twice as long to complete its local computation, allowing us to test our policies under specific, heterogeneous conditions.
For dynamic, real-world scenarios, we use Google's publicly available cluster power traces~\cite{sakalkar2020data} to determine the spare power capacity for each site at any given time. This allows us to simulate realistic power fluctuations where performance is limited by the grid.


\noindent\textbf{Communication delays.}
We synthetically generate the network profiles for our 100 emulated sites where each site is assigned a random geographical distance from the aggregator, ranging from 100 to 10,000 miles (\texttt{distances}). We model the inverse relationship between a site's network bandwidth (\texttt{bandwidths}), which is set to be inversely proportional to its distance, visualized by left y-axis in \autoref{fig:bandwidth_delay_distance}. The total communication delay for each site is then calculated using the~\autoref{eq:communication_per_site_delay}. For the given experiments, we also configure the parameter update size to be 1.5GB to emulate the communication delays associated with large scale ML training.

The right y-axis in  \autoref{fig:bandwidth_delay_distance} illustrates the result of this model, showing how the total communication delay increases non-linearly with distance due to the combined effects of propagation time and decreasing bandwidth.

\noindent\textbf{Evaluation Metric.}
The primary metric for our evaluation is \textit{time-to-accuracy}. This metric measures the total wall-clock time required for a given training configuration to reach a predefined target accuracy on a held-out validation set. We choose this metric because it provides a holistic measure of end-to-end system performance, capturing the impact of both computational throughput and network overhead. A lower time-to-accuracy indicates a more efficient system. We
report results across 10 runs of each experiment.

\subsection{Exploiting heterogeneous power availability}
\label{sec:power_availability_heterogeneous}
Below, we empirically validate the existence of the power-cost trade-off and show the existence of an optimal set of sites which exploit this trade-off to minimize the time-to-accuracy. 

\noindent\textbf{Performance gains with more power availability.}
(\autoref{fig:accuracy_vs_time}) illustrates the direct impact of aggregate computational power on training performance in an idealized setting where all sites have identical power availability and communication costs, thus the communication cost for adding an additional site into the training set is essentially zero. The results clearly show that as more sites are added, the training process yields multiple benefits: the model converges to a higher final accuracy~(\autoref{fig:bar_accuracy}), the time-to-accuracy~(\autoref{fig:bar_time_to_accuracy}) is substantially reduced, and the overall training speedup~(\autoref{fig:bar_speedup}) increases significantly compared to the 10-site baseline. This experiment confirms our foundational premise that, \emph{increasing the total aggregate power by adding more computational units directly improves overall training efficiency} in the ideal case where all the sites reside at a single location thus we can have additional power available without any communication delays. 
Although, we have zero communication cost for adding an additional power site, the returns of having more power available diminish over time in terms of time-to-accuracy. 

Next, we introduce the communication delays based on the communication model discussed previously and discuss the trade-off between power and communication. Here, we observe that the effect of diminishing advantage of adding more sites and increasing communication costs over distances (with lower bandwidths) where selecting more sites into the training set costs more in terms of communication delays, as we discussed in \autoref{sec:design_how_many}.

\smallskip
\noindent
\textbf{Key point.} \emph{We show that in the absence of communication costs, maximizing aggregate power is always beneficial, though with diminishing returns.}

\begin{figure}[t]
    \centering
    \vspace{-0.3cm}
    \includegraphics[width=\linewidth]{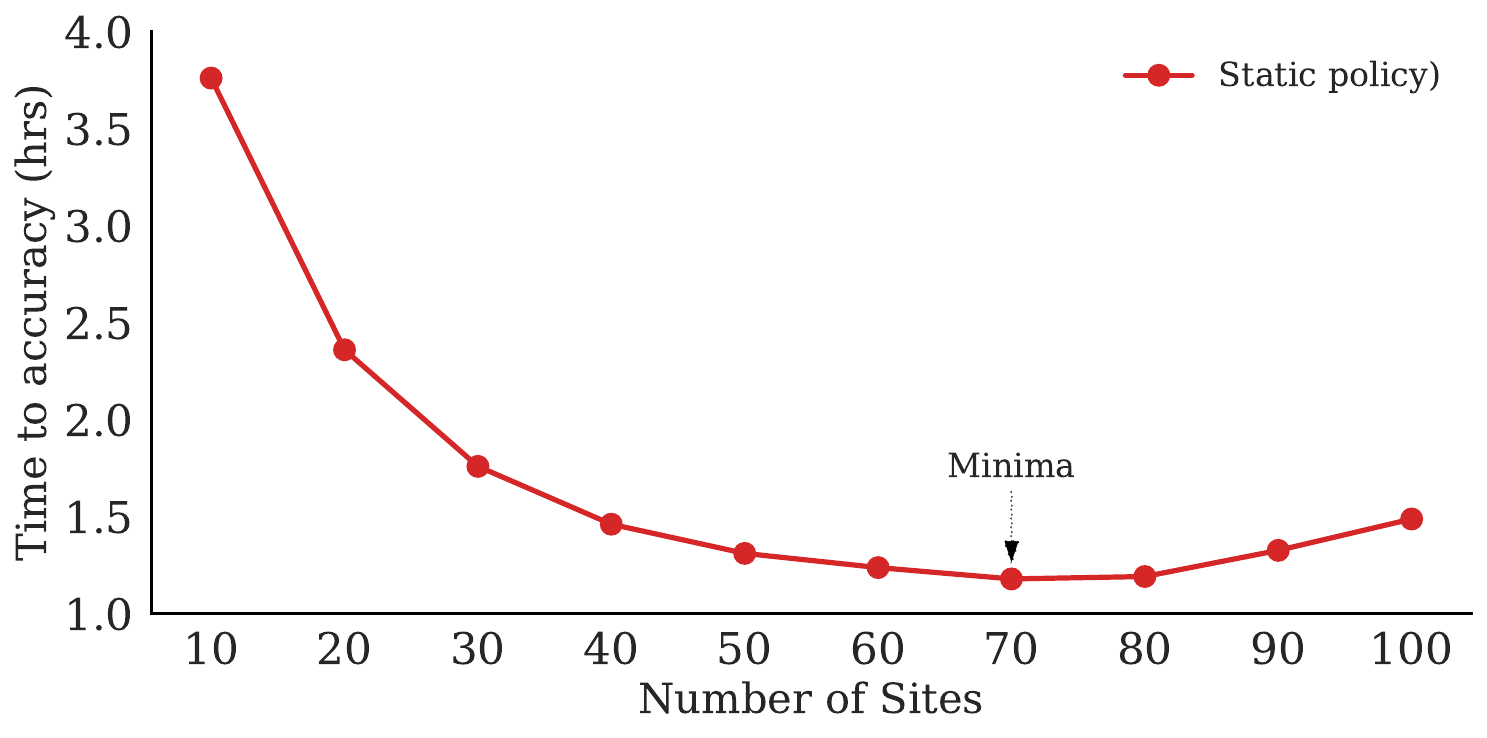}
    \vspace{-0.5cm}
    \caption{\textbf{\emph{The static post-hoc policy identifies an optimal number of sites (in this case, k = 70) that minimizes training time, empirically validating the power-communication trade-off. 
    }}}
    \label{fig:power_comm_tradeoff}
    \vspace{-0.7cm}
\end{figure}

\noindent\textbf{Power-communication trade-off.}
Before evaluating \systemName's performance, we first empirically validate the existence of the power-communication trade-off using the static baseline. For this purpose, we first generate a ranked list of all 100 available sites based on our power-to-cost heuristic, which prioritizes sites with high power availability and low communication delay. We then execute a series of complete, end-to-end training runs on the EMNIST dataset. Each run uses a fixed subset of sites from the top of this list, for instance, using the top $k \in \{10, 20, 30, \dots, 100\}$ sites, the x-axis in ~\autoref{fig:power_comm_tradeoff}. For each value of $k$, we measure the time-to-accuracy where the target accuracy is 45\% in this case, which is the lowest achievable accuracy among all sites.

\autoref{fig:power_comm_tradeoff} explores the trade-off between increased power availability and the communication delays. As we increase the number of sites, $k$, the total time-to-accuracy follows a convex curve. For a small number of sites $k$, adding more sites increases the aggregate computational power, resulting in a reduction in the computational time per synchronization round due to increased power availability far outweighing the increase in communication delay, which leads to a drop in the overall time-to-accuracy. However, as we continue to add sites with relatively low power availability and higher communication delays, the marginal computational gain from each new site decreases. Simultaneously, the probability of including a geographically distant site with high network latency increases. In \autoref{fig:power_comm_tradeoff}, the dashed red line shows that the optimal number of sites occurs at \textbf{\textit{$k^* = 70$}}. After this, the penalty from the increased communication time begins to dominate the diminishing computational benefits, increasing the total time-to-accuracy. Importantly, the minima at $k^*$, is only valid for this specific configuration of the workload and network conditions. Any change would require another expensive, exhaustive search, highlighting the impracticality of a static approach and motivating the need for a robust, dynamic policy that can optimize $k^*$ at runtime.

The existence of this \textit{valley} in the performance curve empirically shows that an optimal number of sites exists for a \textit{given} configuration. While this offline method is too resource-intensive for practical use in a dynamic system, it provides a good benchmark. The performance achieved with $k^*$ sites represents the best outcome for any policy that uses a \textit{fixed set of participants}, and we use this result as a baseline against which we evaluate \systemName's performance.

\smallskip
\noindent
\textbf{Key point.} \emph{We use a static policy to empirically validate the power-communication trade-off, showing that performance degrades after an optimal number of sites is reached because communication costs begin to outweigh computational gains.}

\subsection{\systemName's Dynamic Policy Performance}
\label{sec:dynamic_performance_evaluation}
Having empirically validated the power-communication trade-off and identified the static configuration ($k^*=70$), we now evaluate the performance of our dynamic greedy heuristic policy. As detailed in \autoref{sec:dynamic_policy}, \systemName is designed to find this optimal operating point at runtime, without the need for the exhaustive offline analysis required by the Static Optimal policy.

\begin{figure}[t]
    \centering
    \includegraphics[width=\linewidth]{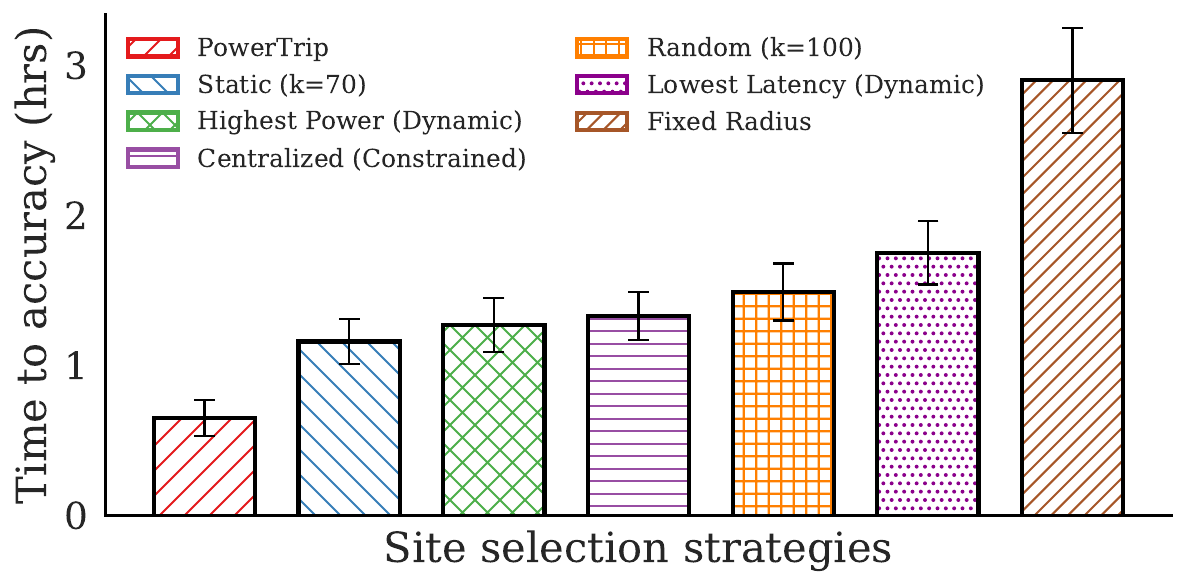}
    \vspace{-0.5cm}
    \caption{\textbf{\emph{Performance evaluation of \systemName's dynamic policy against the baselines. }}}
    \label{fig:evaluation_time_to_accuracy}
    \vspace{-0.5cm}
\end{figure}

To demonstrate this, we conduct an end-to-end training experiment on the EMNIST dataset, comparing the time-to-accuracy of \systemName against the established baselines, including the static optimal policy fixed at $k=70$ and a power constrained centralized training approach. This evaluation has two primary objectives. First, by comparing against the Static Optimal policy, we aim to show that our dynamic approach can achieve better performance without requiring any expensive, offline profiling to pre-configure the system. Second, by comparing against the centralized baseline, we demonstrate that the performance benefit gained from dynamically aggregating heterogeneous power across distributed sites can outweigh the inherent communication overhead.
By monitoring the number of sites \systemName chooses in each round, we can observe its ability to optimize for the number of sites and balance the power-communication trade-off in a practical, online manner.

\autoref{fig:evaluation_time_to_accuracy} presents the results of our evaluation, comparing the time-to-accuracy of each policy. The result clearly demonstrates that \textit{\systemName outperforms all other strategies}, achieving the target accuracy in just \textbf{0.65 hours}. The \textit{Fixed Radius} policy, with a defined radius of 1000 miles (we assume this number based on our communication profiling~\autoref{fig:bandwidth_delay_distance} which shows the bandwidth remains relatively constant in this radius), limited to only 22 nearby sites, is the slowest at 2.9 hours, as it is significantly bottlenecked by its limited access to aggregate power despite its low communication latency. Conversely, the policy that dynamically selects the \textit{Highest Power sites} is also inefficient (1.27 hours), as it ignores communication costs and only includes powerful but distant sites that have high communication delays. Finally,  \textit{Random} selection performs poorly (1.49 hours) due to its inability to make informed decisions about site quality.

Most importantly, \systemName's time of 0.65 hours is nearly \textbf{44\% faster} than the \textit{Static Optimal (k=70)} policy, which took 1.16 hours and is \textbf{50\% faster} than the \textit{centralized approach with 25\% power availability} policy, which took 1.33 hours whereas it performs similar if the power constrained is 50\%. This demonstrates the significant benefit of dynamic adaptation over a fixed, pre-computed strategy and a power-limited centralized approach. 

\begin{figure}[t]
    \centering
    \includegraphics[width=\linewidth]{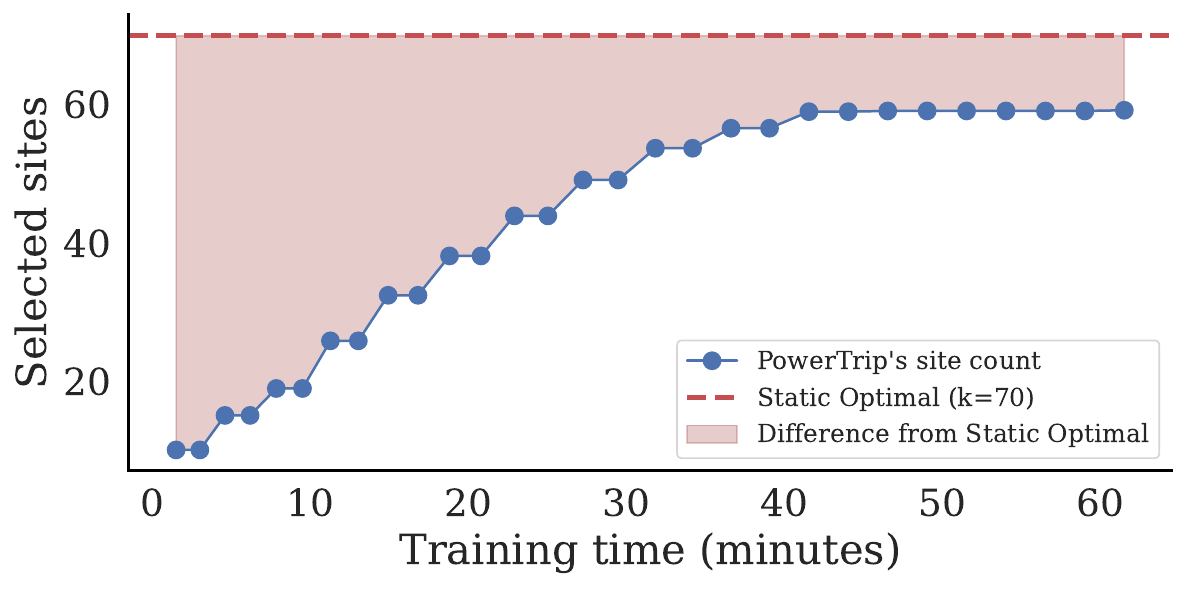}
    \vspace{-0.5cm}
    \caption{\textbf{\emph{Number of sites selected by \systemName's dynamic policy, which scales up over time to avoid high initial costs, against the fixed Static Optimal baseline. }}}
    \label{fig:static_vs_dynamic_site_count}
    \vspace{-0.5cm}
\end{figure}

\noindent\textbf{Advantage in time-to-accuracy.}
\autoref{fig:static_vs_dynamic_site_count} provides a clear visualization of the operational difference between \systemName's dynamic policy and the Static Optimal baseline. The figure plots the number of sites selected by each policy against the total elapsed training time. The Static Optimal policy, by definition, uses a fixed set of $k=70$ sites from the very first round, as indicated by the horizontal dashed line. In contrast, \systemName's policy employs a bottom-up scaling approach. It begins the training process with a small number of highest power, lowest communication latency sites—those offering the best ratio of computational power to communication cost. As the training progresses, it dynamically adds more sites as explained in \autoref{sec:design}.

\begin{figure}[t]
    \centering
    \includegraphics[width=\linewidth]{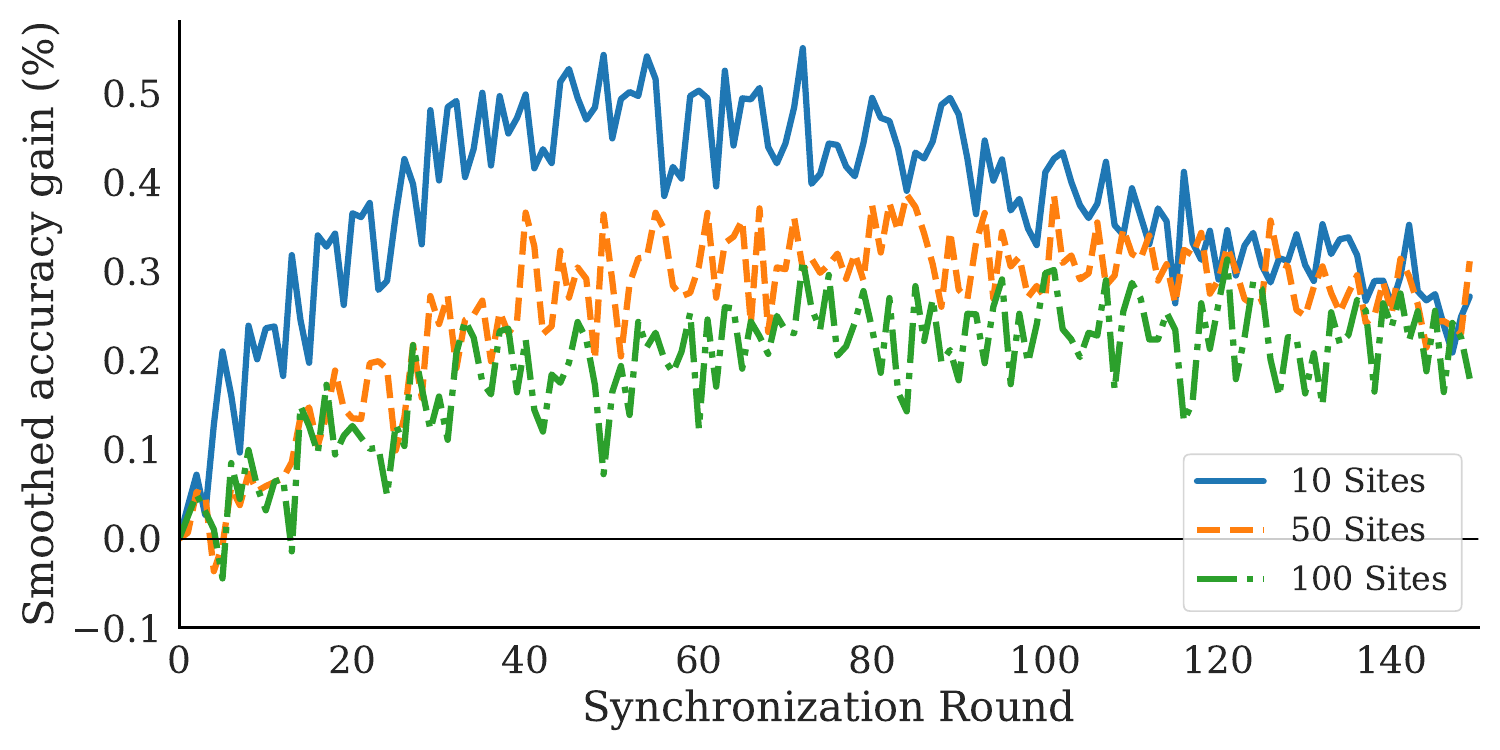}
    \vspace{-0.3cm}
    \caption{\textbf{\emph{Accuracy gain per synchronization round. }}}
    \label{fig:diff_accuracy_gain}
    \vspace{-0.3cm}
\end{figure}

The shaded area between the two curves represents the significant ``site overhead'' incurred by the static policy in the early stages of training. By starting with 70 sites, the static approach is immediately bottlenecked by the highest-latency site within that large group, making every single round slower. \systemName's dynamic policy avoids this penalty. This cumulative time saved during the initial and middle phases of training is the primary reason for its performance improvement over static selection.  To further validate the design of our greedy heuristic, we analyze the marginal benefit of adding more sites in terms of accuracy gain per round. \autoref{fig:diff_accuracy_gain} plots the accuracy gain between larger site configurations (50 and 100 sites) and a small baseline of 10 sites. The key observation is that in the initial phase of training, e.g., the first 10-20 rounds, this difference in accuracy gain is small. While adding more sites does eventually lead to faster convergence, the immediate, per-round improvement in learning is small at the beginning.

This explains the fundamental inefficiency of a static policy. A static approach that starts with a large number of sites (e.g., $k=70$) pays a large and fixed communication cost in every single round, all for a minimal increase in accuracy gain during the early stages of learning. In contrast, a bottom-up greedy approach starts with a small group of efficient, high-power, low-latency sites. In these initial rounds, it achieves the same relative accuracy progress as a large configuration but with significantly lower communication time. The cumulative time saved by avoiding this high, upfront communication tax is the primary reason \systemName achieves a better overall time-to-accuracy, even though it may scale up to a similar number of sites later in the training process. This validates that our online, greedy heuristic effectively navigates the power-communication trade-off by avoiding the over-provisioning of resources when the marginal benefit is low, leading to better performance compared to a static policy.

\begin{figure}[t]
    \centering
    \includegraphics[width=\linewidth]{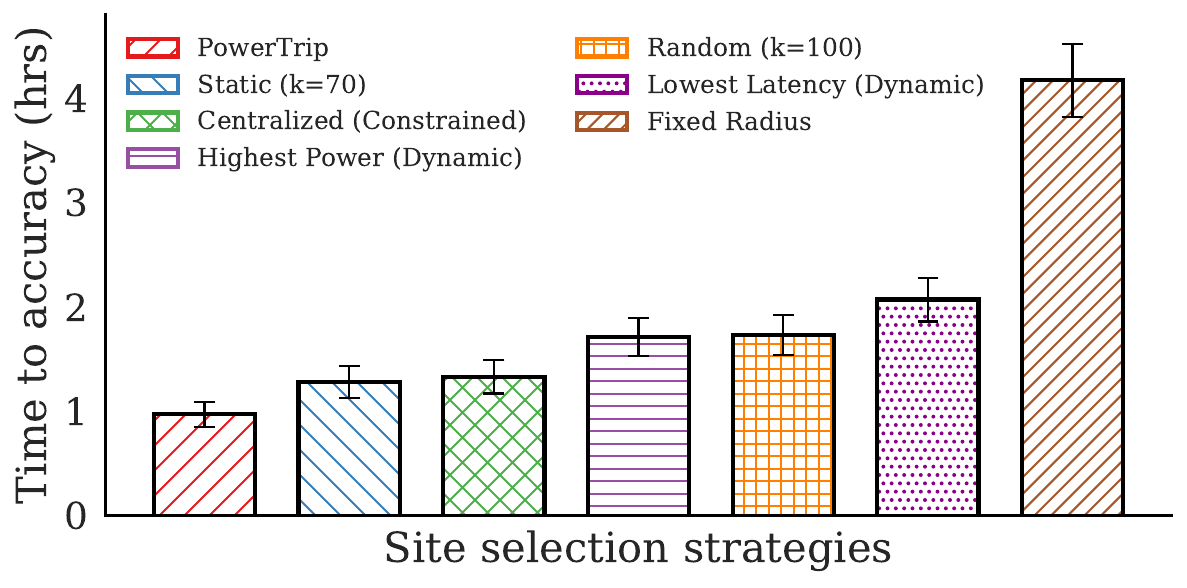}
    \vspace{-0.5cm}
    \caption{\textbf{\emph{Performance evaluation of \systemName's dynamic policy against the baselines using the Google's data center production power utilization traces. }}}
    \label{fig:google_evaluation_time_to_accuracy}
    \vspace{-0.5cm}
\end{figure}

\noindent\textbf{Evaluation with Google Power Traces.} To evaluate the performance of our system under more realistic conditions, we use Google's publicly available data center power utilization traces~\cite{sakalkar2020data}. In this scenario, the power availability at each site is not a synthetic value but is instead determined by the real-time spare capacity derived from the production workload traces. This allows us to model a dynamic environment where a site's computational throughput is limited by its power budget, not just its hardware inventory.

\autoref{fig:google_evaluation_time_to_accuracy} show that the relative performance trends among the policies are consistent with our controlled experiments (~\autoref{fig:evaluation_time_to_accuracy}): \systemName remains the most efficient policy, outperforming all baselines. \systemName's dynamic policy reaches the target accuracy 24\% faster than the optimal static policy and 27\% faster than the centralized approach with 25\% of power availability (which is the lowest power availability from the google power traces). However, a key insight is that the maximum available spare power at any given site in these traces is much lower than in our synthetic distribution, typically falling within the range of 26\% to 51\% of the total capacity. Because the overall power available to the system is more constrained, the absolute time-to-accuracy for \textit{all} policies increases compared to the previous experiments. This result validates the robustness of \systemName’s dynamic approach with the realistic power capacities but also emphasizes the influence of real-world power constraints on overall training performance.

\smallskip
\noindent
\textbf{Key point.} \emph{\systemName outperforms existing baselines by leveraging dynamic scaling to optimize the number of sites at runtime that effectively exploit the power-cost trade-off to minimize time-to-accuracy.  }

\begin{figure}[t]
    \centering
    \includegraphics[width=\linewidth]{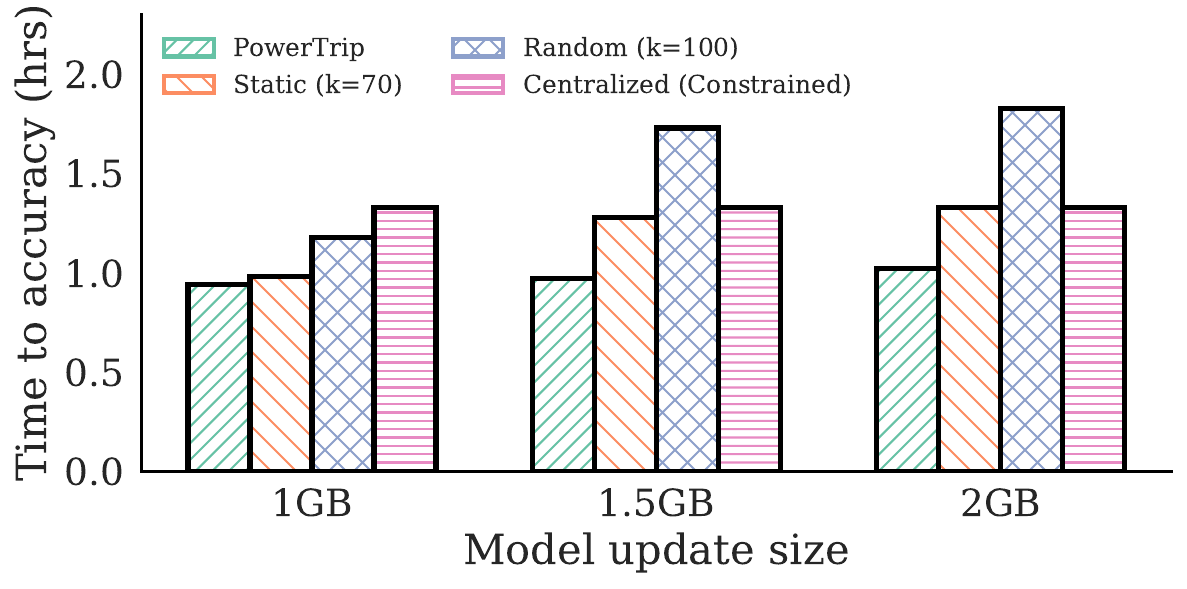}
    \vspace{-0.5cm}
    \caption{\textbf{\emph{Time-to-accuracy comparison across varying model update sizes, demonstrating that \systemName consistently outperforms all baseline policies and highlighting the robustness.}}}
    \label{fig:datasize_comparison}
    \vspace{-0.5cm}
\end{figure}

\subsection{Generalizability and Robustness}
\label{sec:generalization}

To ensure that \systemName's performance benefits are not limited to a single workload, we conduct further experiments to evaluate its generalizability across different datasets and its robustness to varying model update sizes.

\noindent\textbf{Effect of varying parameter update sizes.}
To validate the robustness of our adaptive selection strategy, we conduct experiments using three distinct model's parameter update (that needs to be communicated between the aggregator and the sites) sizes: Small (1GB), Medium (1.5GB), and Large (2GB). As detailed in our design (\autoref{sec:design_which}), \systemName's site selection score is weighted to adaptively prioritize either computational power or low communication cost based on the characteristics of the workload. 

The figure~\autoref{fig:datasize_comparison} shows that \systemName consistently outperforms all baseline policies across all tested update sizes. For the small (1GB) update, local computation speed is more important than the communication delay, thus ($w_p > w_c$) and \systemName achieves a time-to-accuracy of 0.94 hours. As the update size increases to 2GB, the communication delay is the primary bottleneck, the ($w_c > w_p$) and \systemName's time-to-accuracy only modestly increases to 1.02 hours, still giving 23.3\% better time-to-accuracy than the static and centralized baselines (both 1.33 hours). This confirms that \systemName's adaptive selection strategy is effective for a wide range of training workloads.

\begin{figure}[t]
    \centering
    \includegraphics[width=\linewidth]{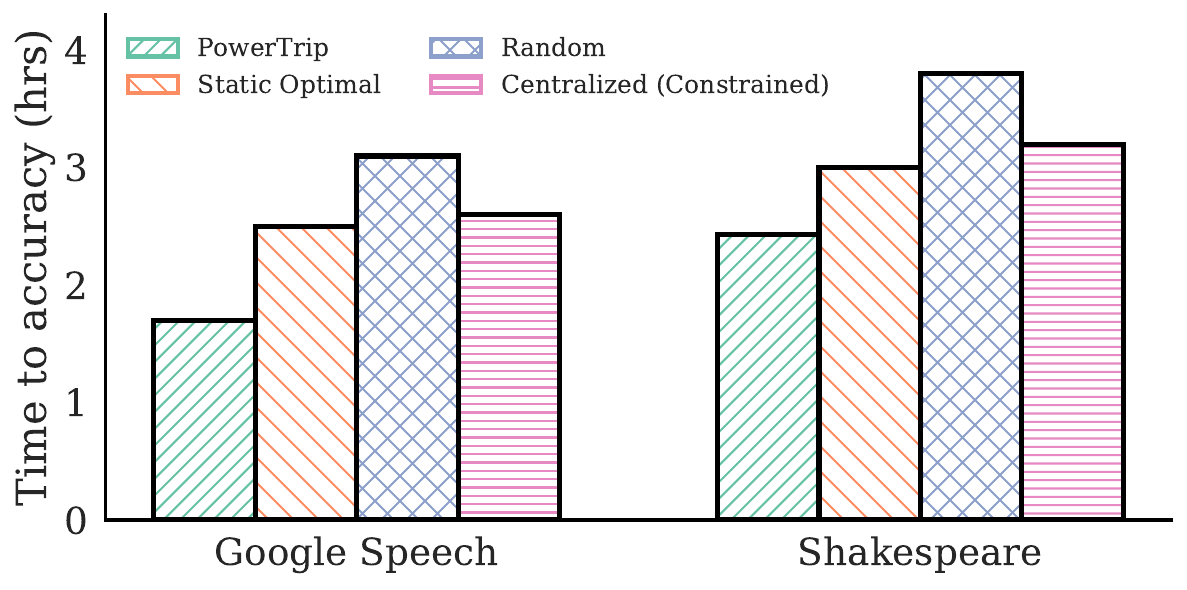}
    \vspace{-0.5cm}
    \caption{\textbf{\emph{Time-to-accuracy comparison across different datasets, demonstrating that \systemName consistently outperforms all baseline policies and highlighting the robustness.}}}
    \label{fig:datasets_time_accuracy}
    \vspace{-0.5cm}
\end{figure}

\noindent\textbf{Performance Across Diverse Datasets.} We repeat our end-to-end training experiments on two additional, widely-used datasets: the Shakespeare dataset for next-character prediction (using an LSTM) and the Google speech commands dataset for keyword spotting (using a KWT-1 model). The results confirm that \systemName consistently performs better across different data modalities and model architectures. As shown in~\autoref{fig:datasets_time_accuracy}, on the Google speech commands task, \systemName achieves the target accuracy 32\% faster than the static optimal policy for that workload. Similarly, on the Shakespeare dataset, our dynamic policy still outperforms the static baseline by 19\%. This demonstrates that our dynamic, greedy heuristic is a robust approach that effectively maximizes the performance across a variety of common ML tasks.

\smallskip
\noindent
\textbf{Key point.} \emph{\systemName's adaptive selection strategy proves effective across a wide range of training workloads, demonstrating its robustness on different datasets and models.}

\section{Related Work}
\label{sec:related}
Our work is positioned at the intersection of distributed machine learning, resource management, and performance optimization. Below, we review prior work to show how \systemName differs. 

\noindent
\textbf{Communication-Efficient Distributed and Federated Learning.}
The challenge of communication overhead in distributed training is well-established~\cite{dean2012large, konecny2016federated, shi2020quantitative}. A significant body of research has focused on reducing the volume of data that must be transmitted during synchronization~\cite{alistarh2017qsgd, konecny2016federated, lin2017deep}. Techniques such as gradient quantization~\cite{alistarh2017qsgd}, where gradients are represented with fewer bits, and sparsification~\cite{aji2017sparse}, where only a subset of significant updates are sent, have proven effective. More recently, parameter-efficient fine-tuning (PEFT) methods like Low-Rank Adaptation (LoRA)~\cite{hu2022lora} and QLoRA~\cite{dettmers2023qlora} have become standard for distributed LLM training. Systems like DiLoCo~\cite{douillard2023diloco} and Photon~\cite{sani2024photon} leverage these PEFT techniques to enable training across many distributed clients with substantially lower bandwidth requirements. Similarly, FusionLLM~\cite{tang2024fusionllm} introduces adaptive compression methods to further optimize communication in decentralized settings.
While these approaches are critical, they primarily optimize \textit{what} is sent, not \textit{who} sends it. They reduce data volume but do not dynamically select participants based on their power, compute, and network resources. In contrast, \systemName introduces a higher-level optimization layer that decides which client sites should participate in the distributed ML training to maximize overall system performance.

\noindent
\textbf{Resource-Constrained AI Training.}
Prior work addresses the growing resource constraints in AI~\cite{patterson2021carbon, shi2024greenllm, singh2024democratizing}. Research on ``Green AI'' has focused on developing models and training methods that are more computationally and energy-efficient~\cite{schwartz2020green}. This includes creating more efficient model architectures and algorithms that require fewer floating-point operations to converge. Prior work schedules training jobs to align with periods of high renewable energy availability or low electricity prices~\cite{wu2022sustainable}.

SkyPilot~\cite{yang2023skypilot}, an inter-cloud broker provisions limited GPU resources across different cloud providers, optimizing for cost and availability. Prior work~\cite{strati2024ml} also analyzes cross-region training as a solution to single-site GPU shortages. The work confirms that while aggregating resources across regions is beneficial, the performance is often bottlenecked by high inter-region network latency, highlighting the same core trade-off we address. These works assume a constant homogeneous power supplies and ignore the challenge of heterogeneous power availability between sites. 

Systems above address resource scarcity but focus on different aspects. Green AI focuses on algorithmic efficiency, while SkyPilot focuses on resource provisioning and cost before a job starts. Prior work hasn’t addressed optimizing distribution to maximize performance gains for large-scale AI training workloads. 

\noindent
\textbf{Dynamic Scaling and Participant Selection Policies~\cite{client_selection_survey1, client_selection_survey2}.}
The concept of dynamic participant selection is well-explored in distributed machine learning. Prior systems like EcoLearn~\cite{EcoLearn2023}, FedCS~\cite{nishio2019client}, and Oort~\cite{lai2021oort} select clients to optimize for specific goals such as carbon footprint, training deadlines, or statistical utility. While effective, these methods typically focus on a single dimension of the problem, such as data quality or energy use~\cite{li2020federated, cho2022flame}. None of this prior work proposes a selection strategy that holistically optimizes for end-to-end training time by dynamically balancing the aggregate computational power of a set of sites against their combined communication cost. Furthermore, many existing scaling policies can be reactive~\cite{EcoLearn2023,clp_accordion}. In contrast, \systemName uses a proactive, bottom-up greedy policy guided by a single objective: maximizing training efficiency in power-constrained environments.

\section{Conclusion}
\label{sec:conclusion}
This paper addresses the problem of optimizing the computational throughput and communication latency in geo-distributed, large-scale AI training to overcome the power and resource limitations of a single site. We introduce \systemName, a system that dynamically optimizes the number of training sites at runtime by implementing a greedy heuristic to identify the point where adding more sites becomes detrimental to performance (time-to-accuracy). Our evaluation demonstrates that by exploiting this trade-off, \systemName can reduce the time-to-accuracy by up to 50\% compared to existing baselines. Future work will involve deploying \systemName on a public cloud to validate its performance in a real-world setting and extending its optimization heuristic to consider the monetary costs of compute instances and data egress.

\balance
\bibliographystyle{plain}
\bibliography{paper}

\end{document}